\documentclass{aa}
\usepackage{graphicx}
\usepackage{longtable,lscape}
\usepackage[varg]{txfonts}

\usepackage{natbib}
\bibpunct{(}{)}{;}{a}{}{,}
\begin{document}
   \title{A young double stellar cluster in a HII region, emerging from its parent molecular cloud \thanks{Based on observations made with the Nordic Optical Telescope,
operated on the island of La Palma jointly by Denmark, Finland,
Iceland, Norway, and Sweden, in the Spanish Observatorio del
Roque de los Muchachos of the Instituto de Astrofisica de Canarias. Based on observations collected at the ESO 8.2-m VLT-UT1 Antu telescope (program 66.C-0015A).}
}


   \author{Jo\~ao L. Yun\inst{1}
		\and
		Anlaug Amanda Djupvik\inst{2}
		\and
          Antonio J. Delgado\inst{3}
	\and
	Emilio J. Alfaro\inst{3}
          }


   \institute{Universidade de Lisboa - Faculdade de Ci\^encias \\
Centro de Astronomia e Astrof\'{\i}sica da Universidade de Lisboa, \\
Observat\'orio Astron\'omico de Lisboa, \\
Tapada da Ajuda, 1349-018 Lisboa, Portugal\\
             \email{yun@oal.ul.pt}
	\and
   Nordic Optical Telescope\\
Apdo 474, 38700 Santa Cruz de La Palma, Spain\\
              \email{akaas@not.iac.es}
	\and
   Instituto de Astrof\'\i sica de Andaluc\'\i a, CSIC, \\
    Apdo 3004, 18080-Granada, Spain\\
              \email{delgado@iaa.es}, \email{emilio@iaa.es}
             }

   \date{Received November 19, 2007; accepted }

 
  \abstract
   {}
   {We investigate the star formation ocurring in the region towards IRAS~07141-0920 contained in the HII region Sh2-294 (S294). We report the discovery\thanks{A new publication (Samal et al. 2007, see references in the text) with a similar analysis of a field in this region appeared when our paper was in the refereeing process.} of a new young double stellar cluster, and we describe its properties.
}
   {High-resolution optical $UBVRI$ and H$\alpha$ images obtained with ALFOSC mounted on the Nordic Optical Telescope (NOT), near-infrared $JHK_S$ images obtained with NOTCam at the Nordic Optical Telescope, and VLT/ISAAC images obtained through the $H_2$ (2.12 $\mu$m) filter 
were used to make photometric and morphological studies of the point sources and the nebula seen towards Sh2-294.
}
   {\rm The optical images reveal an emission nebula with very rich morphological details. The nebula is composed mainly of ultraviolet scattered light and of H$\alpha$ emission with local morphological features conveying an impression of  great dynamical complexity.
Contrasting with the bright parts of the nebula, opaque patches are seen. 
They appear elongated, filamentary, or clumpy.
Our optical photometry confirms that the identification of the illuminator of the nebula is likely to be a B0.5V star located at a distance of about 3.2 kpc.
Our high-resolution near-IR images reveal an embedded cluster, extending for about 2~pc and exhibiting sub-clustering: a denser, more condensed, sub-cluster surrounding the optical high-mass B0.5V illuminator star; and a more embedded, optically invisible, sub-cluster located towards the eastern, dark part of the nebula and including the luminous MSX source G224.1880+01.2407, a massive protostellar candidate that could be the origin of jets and extended features seen at 2.12 $\mu$m.
The double cluster appears to be clearing the remaining molecular material of the parent cloud, creating patches of lower extinction and allowing some of the least reddened members to be detected in the optical images.
We find 12 main-sequence (MS) and 143 pre-main sequence(PMS) members using three different methods: comparison with isochrones in optical colour-magnitude diagrams, detection of near-IR excess, and presence of H$\alpha$ emission. The most massive star fits a 4~Myr post-MS isochrone. The age of the optically selected PMS population is estimated to be 7-8 Myr. The IR-excess population shows sub-clustering on scales as small as 0.23 pc and is probably much younger.
}
   {}


\titlerunning{A young double stellar cluster... }
\authorrunning{Yun, Djupvik, Delgado, \& Alfaro}

   \maketitle
%

\section{Introduction}

Galactic nebulae are often times clear signposts of regions of star formation due to the proximity of the young stars to the parent material of the molecular cloud that originated the stars.
The presence of nebulosity has been used to find and identify star formation regions \citep[e.g.][]{yun94,tapia97,rubio98}.
In addition, young stars are the source of energetic winds, jets or outflows, whose interaction with their surrounding cloud material produces shocks, which are denounced by nebulae with characteristic shapes (e.g. bow shocks, HH objects \citep[e.g.][]{reipurth01}), and ultimately leads to the clearing of the cloud material and emergence of the young stars out of the cloud core.

The morphology of nebulae seen towards star forming regions can reveal the processes that have been ocurring in the interior of the parent cloud. Diffuse reflection nebulae, both in the optical and in the near-infrared are the result of illumination from bright sources more or less embedded in the cloud. Optical emission nebulae corresponding to excited gas can be detected using narrow-line filters, the most common one being H$\alpha$ \citep[e.g.][]{fich90}.

Small groups or isolated young low-mass stars have limited dynamical effect on their surroundings. On the other hand, rich clusters of young stars or high-mass young stars will often produce spectacular nebulae.
Massive star formation is currently being addressed using a variety of the new instruments made available in the last years. Unlike the low-mass family of young stars, the relatively short timescale for their accretion and pre-main sequence evolution implies that massive stars reach the main-sequence still embedded in their parent molecular cloud cores. On the other hand, their strong hard radiation ionizes their surroundings creating a ultra-compact H~II (UCHII) region, which may represent the earliest recognizable stage in the life of massive stars \citep{churchwell02}.
UCHII regions have been identified using H$\alpha$ surveys \citep[e.g.][]{rodgers60}, radio surveys, and looking for characteristic IRAS colours \citep[e.g.][]{wood89}. More recently, they have been studied using a combination of near and thermal infrared surveys and K-band spectroscopy \citep{bik03,bik06}. 

IRAS~07141-0920 is an IRAS source with colours close to those of ultra-compact HII regions \citep{wood89}. It is seen towards a large, extended optical nebula. 
\cite{rodgers60} detected H$\alpha$ emission, and \cite{fich90} detected the H$\alpha$ line at v(LSR) of 33.5 km~s$^{-1}$.
The source has been first cataloged as an HII region by \cite{sharpless59}
in his catalogue of HII regions (source Sh2-294 or S294).
Using UBV photoelectric photometry, \cite{moffat79} derived a photometric spectral type of B0.5V for the hottest star in this HII region, which they assume to be the ionizing source. Their estimated distance was of 4.6 kpc. 
Extended radio continuum emission from S294 at 1.46 GHz has been detected and mapped, resulting in a diameter for the HII region of at least $7 \times 7$ arcmin$^2$ \citep{fich93}. 
A strong CO $J=1-0$ molecular line transition towards IRAS~07141-0920 has also been detected, at a radial velocity v(LSR) of 32 km~s$^{-1}$ \citep{blitz82,wouterloot89}. They derived a kinematic distance of 3 kpc, much smaller than the value of 4.6 $\pm$ 1.5 kpc derived by optical photoelectric photometry \citep{moffat79}. \cite{wouterloot89} also note that the CO line is flat-topped (non-gaussian), a hint that the molecular gas is not in a quiescent state, possibly with turbulence or other line broadening mechanism present. Because they have a single observation and have not mapped the cloud, there is no estimate of the extent of this molecular cloud seen towards IRAS~07141-0920. 
Searches for masers resulted in non-detections (water masers: \cite{henkel86}; methanol masers: \cite{macleod98}, \cite{codella95}). \cite{bica03} reported the presence of a 2MASS infrared, loose star cluster towards Sh2-294.
 \cite{samal07} provides a multivalength analysis of the Sh2-294 region describing the distribution and nature of the young stellar objects, and the morphology of the thermal ionized
gas and dust emissions\footnote{This article was published while our manuscript was in the refereeing process.}.
Here we present an optical and near-infrared study of the extended optical nebula and the embedded stellar population seen towards Sh2-294 (S294) / IRAS~07141-0920. 
We confirm the presence of a young embedded cluster which appears to be actively in the process of destroying its craddle, that is dissipating the surrounding parent molecular cloud. Most cluster members are still embedded and are either not seen or are very faint on our deep optical CCD images. 
 
Section~2 describes the observations and data reduction. In Sect.~3, we describe the properties of the extended nebula. Section~4 describes the stellar density distribution and the presence of a young embedded cluster. In Sect.~5, we present our photometric results, estimate the reddening, distance, and age of the cluster, and make cluster membership assignments. 
In Section~6, we discuss the spatial distribution of the cluster members and present a possible star formation scenario for this region. A summary is given in section~7.


\section{Observations and data reduction}

\subsection{Optical observations}

Optical ($UBVRI$ and H$\alpha$) observations were carried out during the night of 2006 November 29 with the NOT (Nordic Optical Telescope), operating at the ORM (La Palma) under very good seeing conditions (0.6 arcsec). The ALFOSC (Andalucia Faint Object Spectrograph and Camera) CCD camera in imaging mode was used (with a plate scale of 0.19 arcsec/pixel). 
Short and long exposure images were obtained through each band: short exposures with integration times of 10, 10, 5, 5, 5, and 60\,s for $UBVRI$ and H$\alpha$ (FWHM=3.3 nm) bands, respectively; and long exposures with integration times of 600, 300, 300, 300, 300, and 1200\,s, for the same bands, respectively. 

The observations were reduced with the adequate routines in the 
IRAF\footnote{The Image Reduction and Analysis Facility (IRAF) is distributed 
by the national Optical Astronomical Observatory, which is operated by the 
Association of Universities for Research in Astronomy, Inc. (AURA) under 
cooperative agreement with the National Science Foundation.} package. 
In each of the $UBVRI$-bands, three long exposures were co-added into a single image. Aperture and PSF photometry was performed in both the short and the long exposure images. The resulting magnitudes were then joined together to produce a final list of instrumental magnitudes.

Frames of a standard field were also secured at 
different airmasses. The field includes 12 stars in the SA98, around star 652 
\citep{landolt92}. These observations were used to transform the instrumental 
magnitudes to the standard system, with the procedure described by \cite{delgado98}, based upon the assumption of multilinear dependence of standard 
magnitudes on airmass, instrumental magnitude, and an appropriate colour index.

\subsection{Near-infrared observations}

\subsubsection{NOTCam $JHK_S$ observations}

Near-infrared ($J$, $H$ and $K_S$) images were obtained on 2007 January 
29 using the Nordic Optical Telescope near-IR Camera/spectrograph 
(NOTCam)\footnote{See URL http://www.not.iac.es:/instruments/NOTCam/ for 
more information on NOTCam.}. The detector was the 1024 x 1024 x 18 micron 
Hawaii engineering grade array. The wide field camera ($0.234''$/pix) was 
used, and the observations were performed with a ramp-sampling readout 
mode. Every sky position was integrated for 30 sec, reading the array 
every 5 sec and using the linear regression result of these 6 readouts. 
For each filter, 9 dithered sky positions were observed. Because of varying 
sky conditions, only the best frames were selected, resulting in total 
on-source integration times of 210, 180, and 150 seconds for $J$, $H$, 
and $K_S$, respectively. 

The images were reduced using a set of own NOTCam scripts to correct for 
bad pixels, subtract the sky background, and flatfield the images. 
Differential twilight flats taken a few nights apart were used to correct
for the pixel to pixel variations of the response. The selected images were 
then aligned, shifted, trimmed, and median combined to produce a final $JHK_S$ 
image of $3.7' \times 3.4'$ size. 

Point sources were extracted using {\tt daofind} with a detection threshold of 
6$\sigma$, and a few additional sources were added in by hand. Aperture 
photometry was made with a small aperture (radius = 4 pix, which is about
the measured FWHM of the point spread function) and aperture corrections, 
found from 5 bright and isolated stars in each image, were used to correct 
for the flux lost in the wings of the PSF. The error in the determination 
of the aperture correction was $<$ 0.02 mag in all cases. These errors are added to the value MERR which is output by the IRAF task {\em phot}. 
A total of 491, 496, and 494 sources were found to have fluxes in $J$, 
$H$, and $K_S$, respectively, and errors $\sigma_{Ks} <$ 0.35 mag in the $K_S$ band (the noisiest and least sensitive). All these are included in the table of photometric results for completeness (Table~1, see below), but we include only the 444 NOTCam sources with $\sigma_{Ks} <$ 0.2 mag in our analysis in this paper.

The brightest star in the field S294-B0.5 (ID 717) had peak pixel values outside the linear range of the detector both in the $H$ and $K_S$ bands. 
We therefore used only the first of the 6 ramp-sampling readouts, with 
an effective integration time of 5 seconds for this source. 

We have used the 2MASS All-Sky Release Point Source Catalogue \citep{skrutskie06,cutri03} to calibrate our observations both in terms of photometry and 
astrometry. The plate solution was found using 27 isolated and bright 
2MASS stars over the FOV and the iraf task {\tt ccmap} with the ``tnx'' 
projection allowing for geometric distortion and a 3rd order polynomial 
with full cross terms. NOTCam XY positions were then transformed to RA 
and DEC, and all 2MASS sources were matched to all NOTCam sources using 
a search radius of $0.36''$, which corresponds to three times the positional 
accuracy from the plate solution. We excluded all 2MASS sources that do 
not have photometric quality flag from A to D in all bands. Also a few 
multiples and one clearly variable star were excluded and we were left 
with a total of 85 sources which could be compared in terms of fluxes 
and positions.
The $JHK_S$ zeropoints were determined using 2MASS stars brighter than $K_S$ = 13.5 mag after excluding 2 cases of possible variability (i.e. 17 stars). 
The standard deviation of the offset between NOTCam and 2MASS photometry
is 0.07, 0.05 and 0.05 mag in $J$, $H$, and $K_S$, respectively, for 
sources with $K_S <$ 13.5 mag. We estimate the completeness limit of the observations to be roughly 18.5, 18.0, and 17.5 magnitudes in $J$, $H$, 
and $K_S$, respectively.

\subsubsection{ISAAC $H_2$ observations}

Near-infrared observations, using the H$_2$ filter centred at the $v=1-0$ line at 2.12$\mu$m, were conducted on
2000 November 10 using the ESO Antu (VLT Unit 1) telescope equipped with the short-wavelength arm (Hawaii Rockwell) of the ISAAC instrument. The ISAAC camera \citep{moorwood98} contains a 1024 $\times$ 1024 pixel near-infrared array and was used at a plate scale of 0.147 arcsec/pixel resulting in a field of view of 2.5~$\times$~2.5 arcmin$^2$ on the sky.  At each of 5 different jitter positions (with the length of the jitter box equals to about half the field of view), one image with 60~s exposure time were taken. 

Standard procedures for near-infrared image reduction were applied (e.g. \cite{yun94}) resulting in a final mosaic image for each band. A sky image was computed for each dithered frame. Each frame was then sky subtracted and flat-fielded using master sky flats. The final clean images were corrected for bad pixels while constructing the final mosaic which covers about $5 \times 5$ arcmin$^2$ on the sky. Due to the presence of significant field distortion in these ISAAC images (two pixels at the edges and 2.5 pixels in the corners), the individual frames were corrected before they were put together to form the mosaic. The correction for field distortion was performed using IRAF/GEOTRAN together with the adequate correction files provided by ESO at their web page. The central, co-added region, with enhanced signal-to-noise ratio covers about 2.5~$\times$~2.5 arcmin$^2$ on the sky.

\section{Extended nebula}

\subsection{Morphology and colours}

\begin{figure}
   \centering
   \includegraphics[width=8.5cm, angle=0]{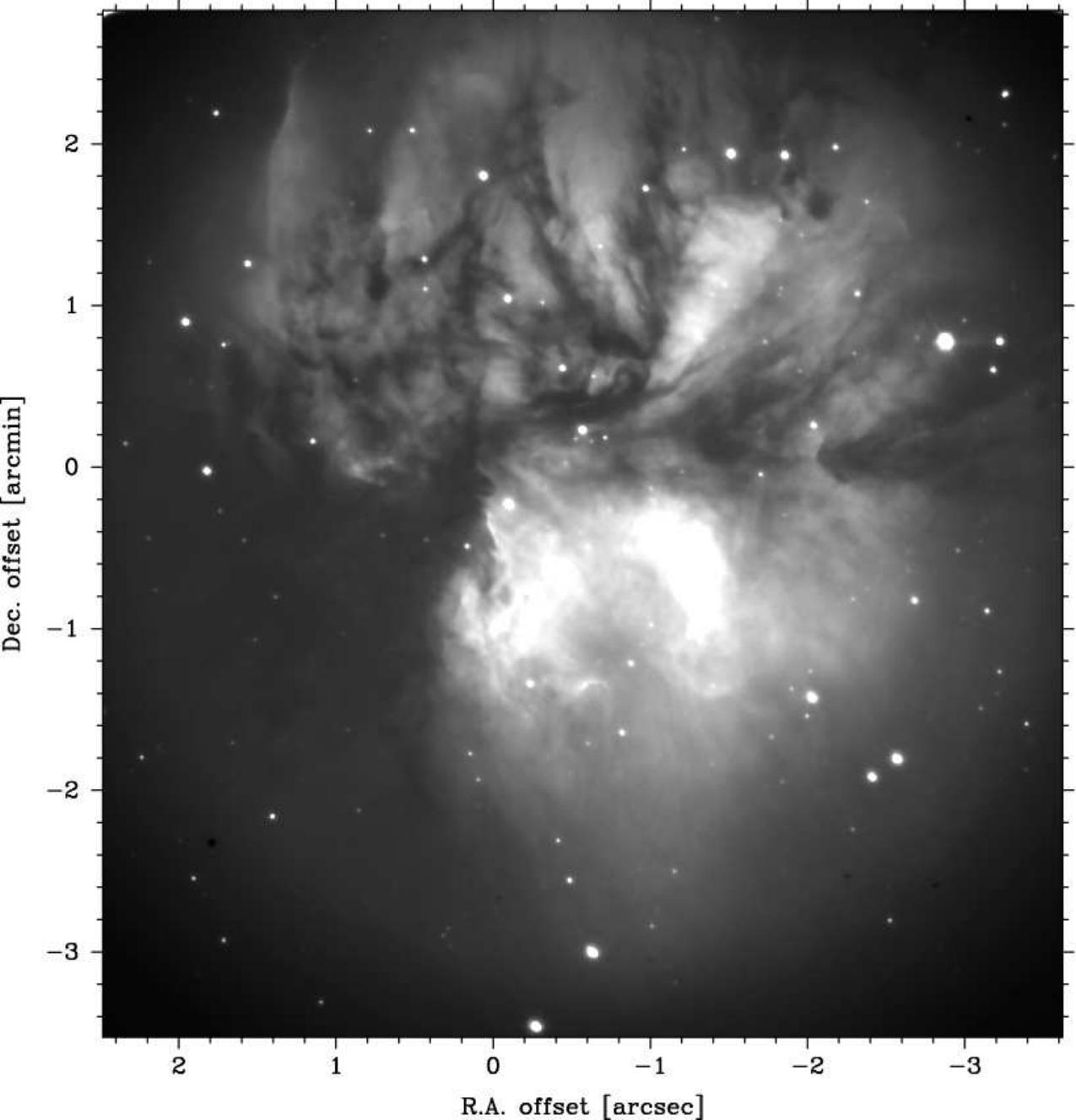}
   \caption{
H$\alpha$ image towards IRAS~07141-0920 covering about $6 \times 6$ arcmin$^2$ ($5.4 \times 5.4$ pc$^2$, for our estimated distance of 3.2 kpc). Notice the bright nebula and the patchy and wispy dark cloud material.
	} 
	\label{fig1}%
    \end{figure}

Figure~\ref{fig1} shows the H$\alpha$ image centered at the IRAS~07141-0920 position and covering about $6 \times 6$ arcmin$^2$.
The high-resolution of ALFOSC allied to the very good seeing during the observation produced an image containing a wealth of detail.
It reveals a large, extended nebula which appears to be illuminated by one or more objects possibly seen in this H$\alpha$ image. 

Contrasting with the bright parts of the nebula, dusty, opaque patches are seen, specially in the northern and eastern sides. These dark clouds appear to be in the process of being shredded or blown apart by stellar winds, jets or outflows phenomena. They appear elongated, filamentary, wispy, clumpy or knotty. The overall image conveys an impression of streams, flows, turbulence, cavities, a region of great dynamical complexity.

%

   \begin{figure}
  \centering
   \includegraphics[width=8.5cm]{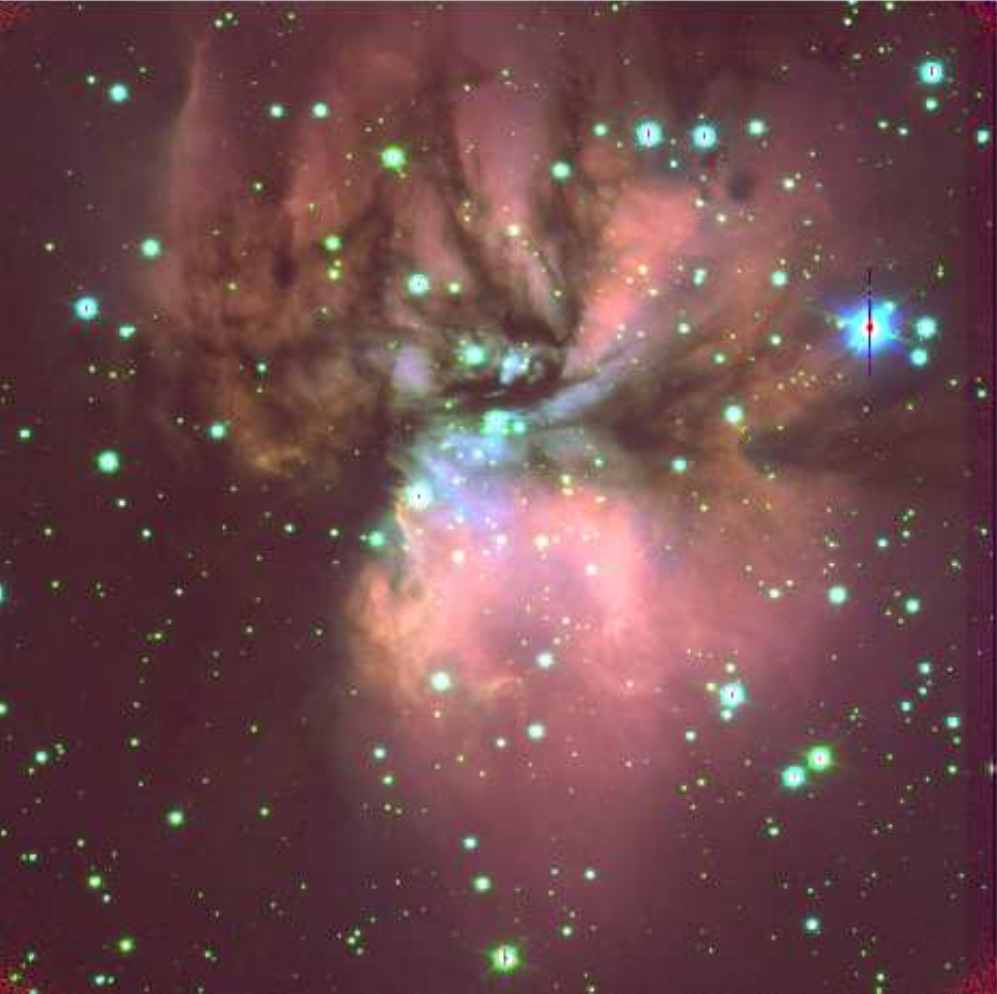}
   \caption{U-band (blue), R-band (green), and H$\alpha$ (red) colour composite image towards IRAS~07141-0920, covering about $6\times 6$ arcmin$^2$. North is up and East to the left. Blue extended emission traces scattered light. Red extended emission traces H$\alpha$ nebular emission from excited gas. The green colour (R-band continuum) appears essentially tracing point sources.	
} 
	\label{fig2}%
    \end{figure}

The main contributions to the nebula brightness are continuum ultraviolet scattered light (mostly contained in the U-band) and H$\alpha$ emission. This is shown in the 
colour composite image shown in Fig.~\ref{fig2}. 
This Figure shows a U-band (blue), R-band (green), and H$\alpha$ (red) colour composite image towards IRAS~07141-0920. With this colour code, the blue extended emission traces scattered light while red extended emission traces H$\alpha$ nebular emission from excited gas. The green colour (R-band continuum) appears essentially tracing point sources. Much of the nebular light seen is composed of H$\alpha$ emission. The contribution of continuum scattered light is stronger near the centre of the nebula where the illuminator is most likely to reside.

\subsection{The illuminator star}

Given the presence of the HII region and the large, extended nebula, at least one O or B star is responsible for the ionization and excitation of the nebular gas. 
In order to identify the illuminator(s), we note that it should be located close to the central region of the nebula which happens to be also where the blue scattered light is brightest.

Because dark lanes of obscuring material can be seen in silhouette against the bright nebula (mostly in the northern part of the nebula), we infer that the illuminator (or illuminators) must be partially embedded in the cloud. The illuminator should not be too deeply embedded, with its light escaping from ``holes'' (regions of lower optical depth, lower extinction) after scattering off the walls of cavities in the near side of the cloud and exciting or ionizing the gas at the surface of the cloud.

\begin{figure}
   \centering
   \includegraphics[width=8.5cm]{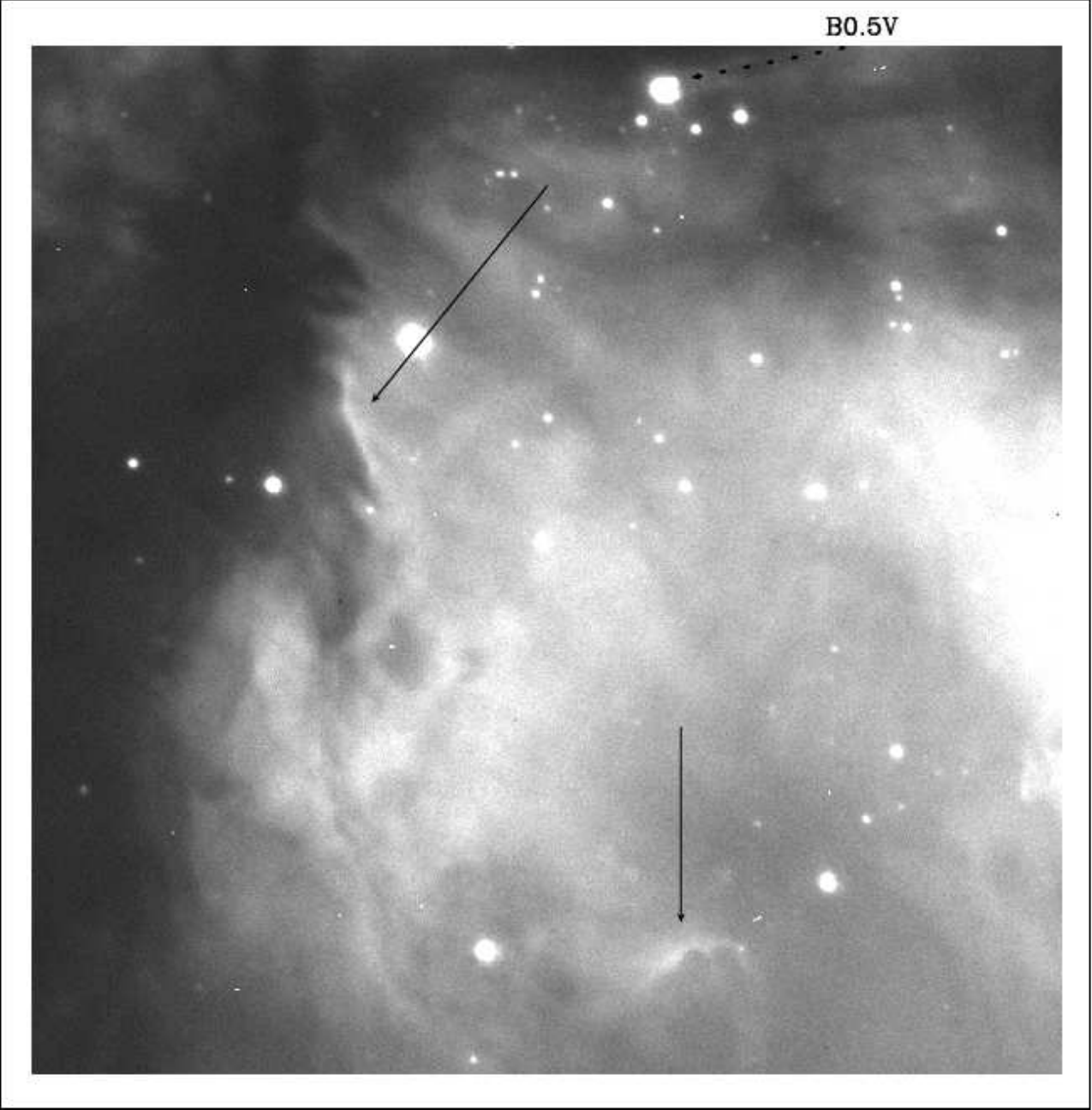}
   \caption{ Zoomed-in H$\alpha$ image of the nebula covering $1.9 \times 1.9$ arcmin$^2$ of the southern and central region of the nebula. B0.5V is the illuminator of the nebula. Notice the rim-bright features, some of them indicated by arrows. North is up and East to the left.
	} 
	\label{fig3}%
    \end{figure}

In this central region, the brightest source is the star marked with the letter ``B0.5V'' in Fig.~\ref{fig3} and we identify this object as being the illuminator of this nebula. This object coincides with the star No. 4 of \cite{moffat79} who first identified it as the hottest star in the region and thus the responsible for the HII region. This identification has been confirmed by \cite{samal07}. In the remaining, we will refer to this object as S294-B0.5V.

Additional arguments support this conclusion. Figure~\ref{fig3} reveals nebular structures that display illuminated rims and shadows implying that the incident light is along the directions indicated by the two solid line arrows. In all cases, the only bright object located within the bright scattered light region is object S294-B0.5V.
In section 5, we discuss optical and infrared photometry results for this object, and the parameters of the associated cluster.

\section{Stellar density distribution: a young stellar cluster}

Together with this massive B0.5V star, this region hosts the birth of a young stellar cluster associated to the massive star.

\begin{figure}
   \centering
   \includegraphics[width=8.5cm]{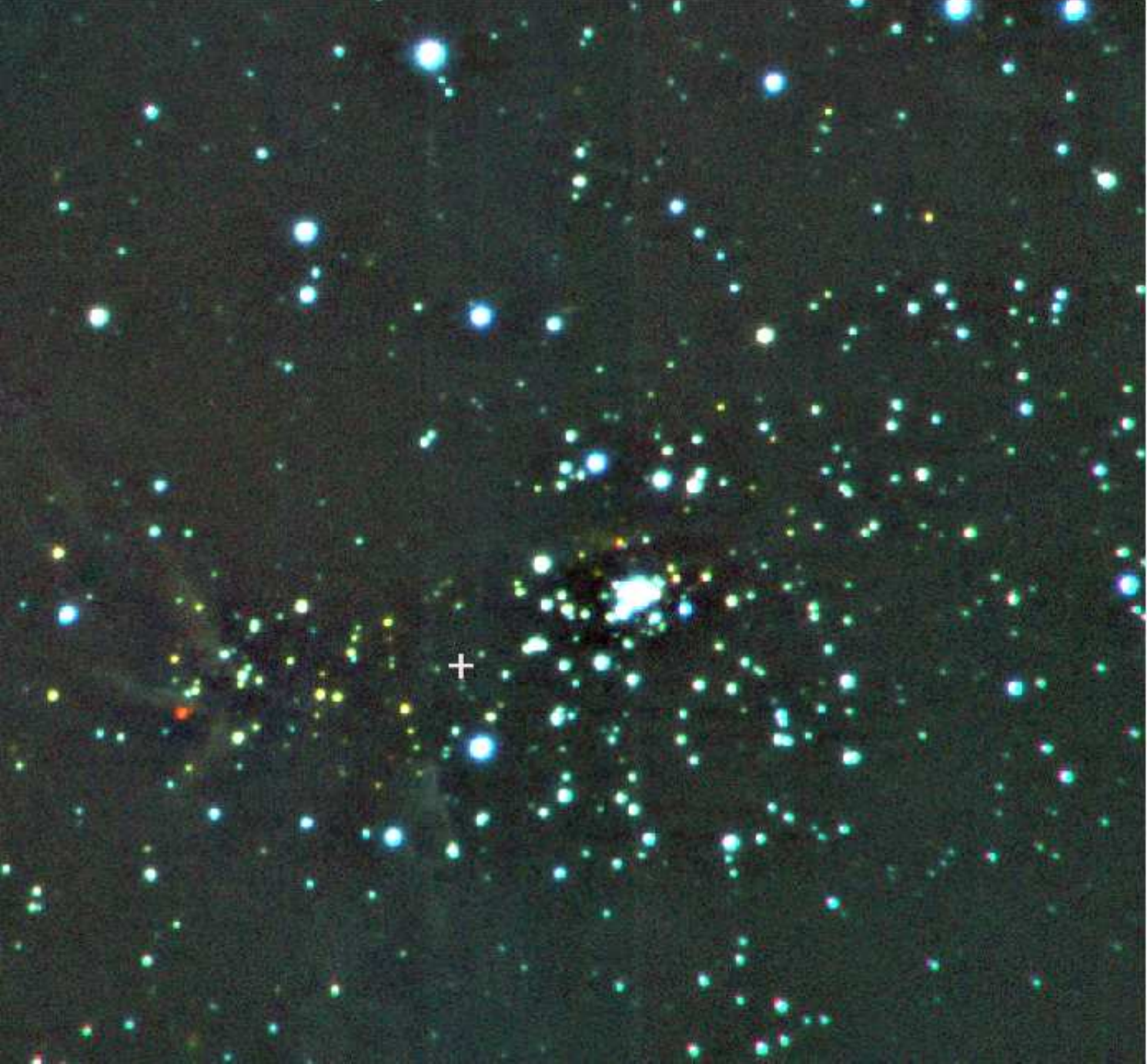}
   \caption{
J (blue), H (green), and Ks (red) colour composite image towards IRAS~07141-0920. The image covers about $3.2\times 3.4$ arcmin$^4$ (dashed-line box in Fig.~7). North is up and East to the left. The white cross indicate the position of the IRAS source ((0, 0) position of Fig.~\ref{fig1})
	} 
	\label{fig4}%
    \end{figure}

Figure~\ref{fig4} presents the NOTCam $JHK_S$ colour composite image obtained towards IRAS~07141-0920. The optical nebula which dominated much of the $UBVR$ and $H\alpha$ images is absent. Instead, a large number of stars are seen, located in the region of the optical nebula but also to the east, where a dark patch of extinction is present in the optical images. 

\begin{figure}
   \centering
   \includegraphics[width=8.5cm]{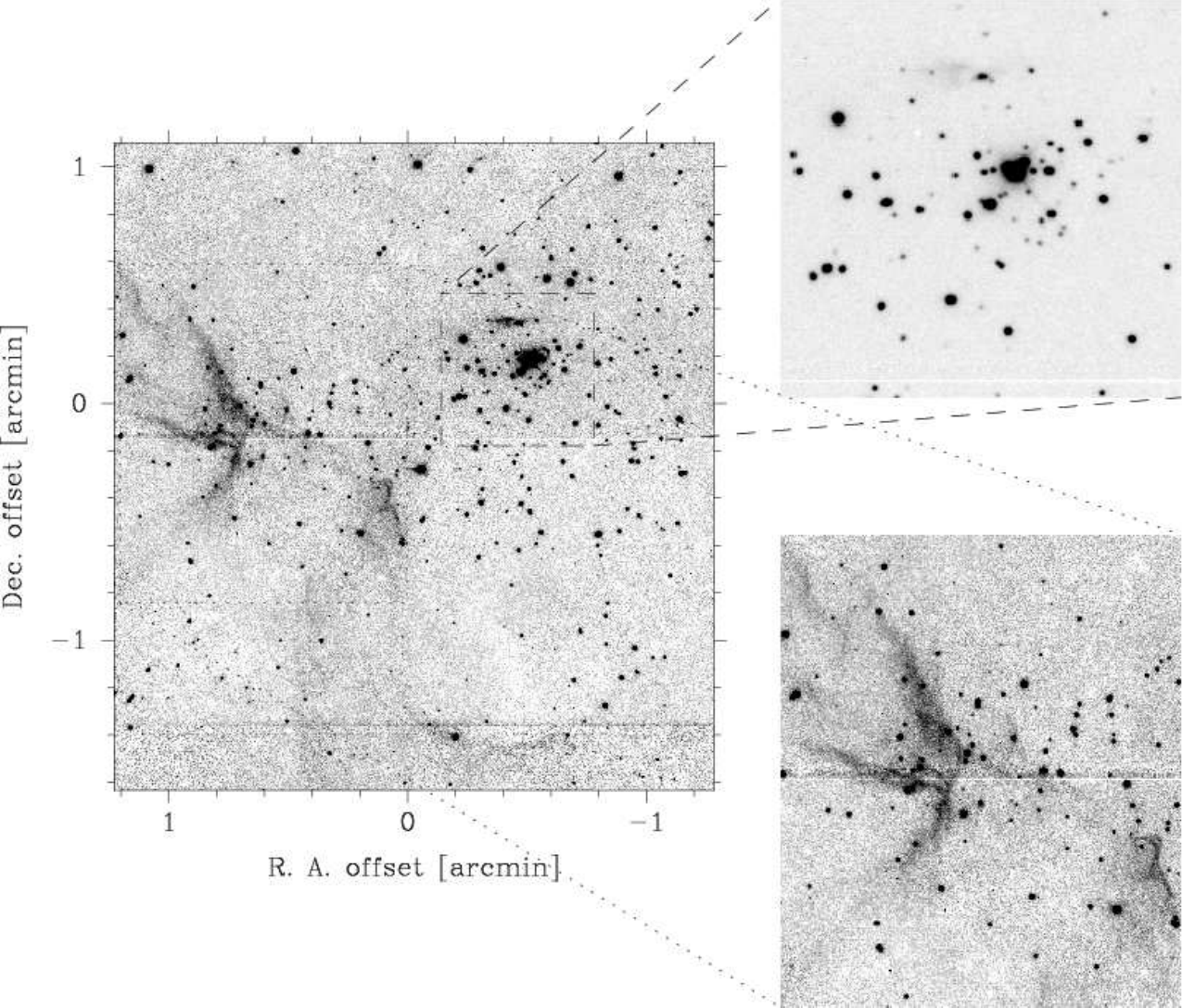}
   \caption{
Central region of the $H_2$ mosaic image towards IRAS~07141-0920. The axes give the coordinates relative to the IRAS source. The two zoomed-in views reveal elongated emission in the eastern part, and the enhanced concentration of stars around S294-B0.5V.
	} 
	\label{fig5}%
    \end{figure}

A higher angular resolution image is provided by the VLT ISAAC $H_2$ image seen in Fig.~\ref{fig5}. 
The stars appear to form two distinct groups or sub-clusters: a western group (the western sub-cluster), that includes S294-B0.5V, less extincted and located towards the optical nebula; and an eastern group (the eastern sub-cluster), redder and located towards a dark optical region.

The eastern part of Fig.~\ref{fig5} also shows the presence of elongated, jet-like or filamentary emission which suggest flows or material being blown away, likely by powerful stellar winds, or affected by ionization fronts.

%

\begin{figure}
   \centering
   \includegraphics[width=8.5cm]{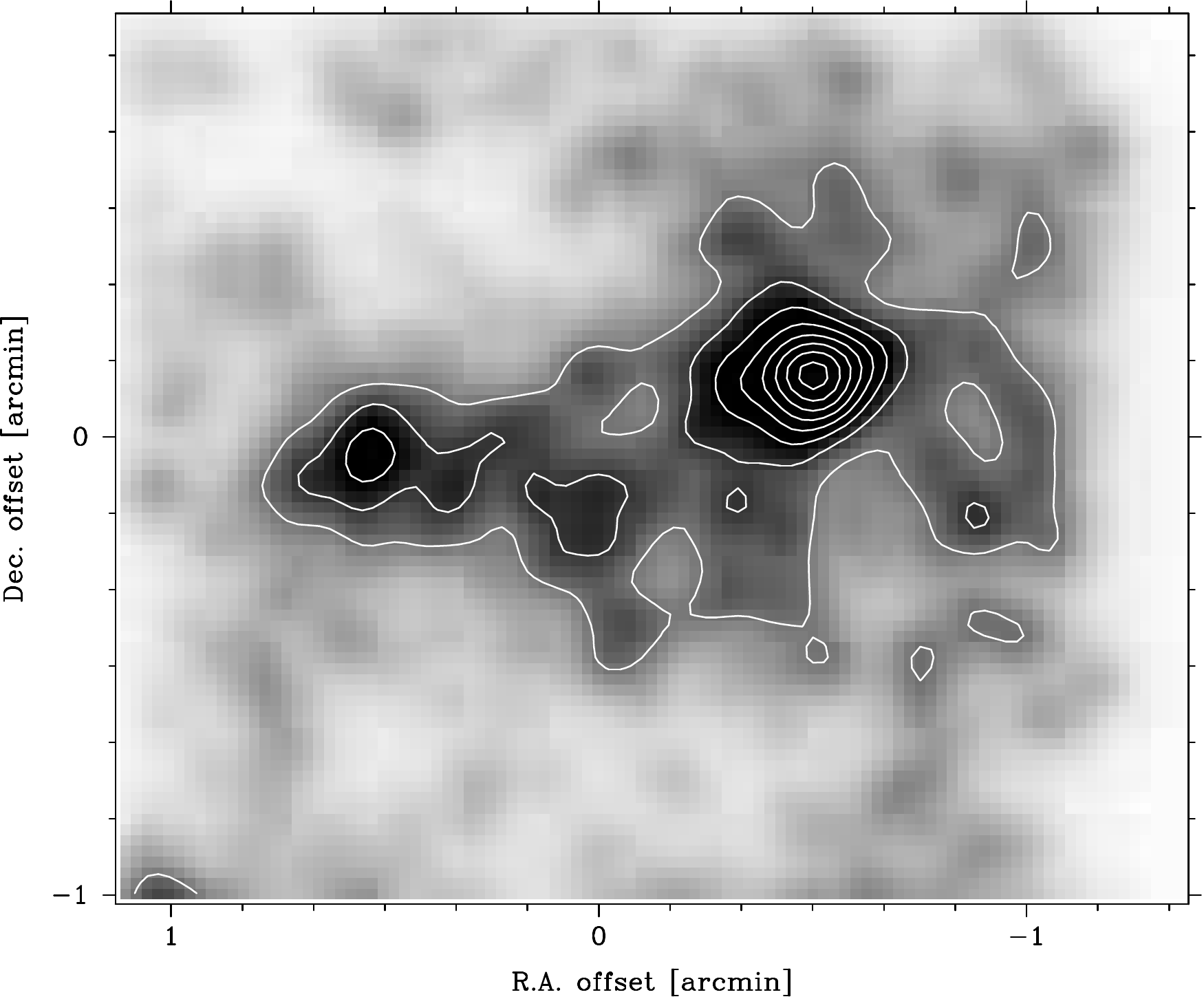}
   \caption{
Contour plot of the stellar surface density of the H$_2$ image. 
Two peaks of the stellar surface density can be identified: the strongest, in the West, is the ``western peak'' ( where the illuminator is located, see text); the second one, in the East, is the ``eastern peak". The axes give the coordinates relative to the IRAS source. The lowest contour level is 9 sigma above the sky background stellar surface density. The highest contour level is about 14 times the value of the sky background stellar surface density.
	} 
	\label{fig6}%
    \end{figure}

The presence of the two groups of stars is more clearly seen in the contour plot of the stellar surface density, shown in Fig.~\ref{fig6}, obtained from the $H_2$ mosaic image
Two peaks of the stellar surface density can be identified. The stronger peak is located in the central region of the optical nebula and we name it the western stellar density peak. It contains S294-B0.5V, the illuminator of the optical nebula, and a steep slope in the number of stars surrounding it. Thus, in a small region around the massive S294-B0.5V star, there is a large number of young lower-mass stars (see also the zoomed-in upper-right panel of Fig.~\ref{fig5}). The stellar surface density at this western peak is about 14 times higher than the value of the sky background stellar surface density.
The second peak is located in the dark region to the east of the nebula and we name it the eastern stellar density peak. Together, the two peaks correspond to the presence of a young stellar cluster showing sub-structuring in the scale of about 2 pc, the size of the region.  
The position of the IRAS source, at offset (0, 0), is located between the two peaks with the IRAS uncertainty ellipse along the east-west direction. 

Interestingly, the map of 1.46 GHz continuum emission \citep{fich93} also displays two peaks located approximately at the same positions of the peaks of the stellar surface density. However, the stronger radio continuum peak is associated with the eastern peak, that is with the more embedded sub-cluster, invisible in the optical images.
The 1280 MHz radio continuum map of \cite{samal07} contains a peak coincident with the stronger 1.46 map, and thus the eastern stellar density peak.

%

\section{Photometric results}

\subsection{Standard calibration} 

As mentioned above, observations of standards stars in the Landolt catalogue \citep{landolt92} were secured for standard optical calibration purposes. On the other hand, when additional reliable standard values are available for a region of the sky, an advantageous calibration procedure is at hand, by means of direct correlation between the instrumental values and those published standard values \citep{delgado00,delgado07}. The only published photometric study including stars of this region (\cite{moffat79}; MFJ in the following) contains $UBV$ indices for 15 stars included in the field of our $UBVRI$ observations.
Of these 15 stars, we have good, non-saturated photometry for 14 stars. The correlation with the values from this study can be performed with some caution. Several stars included in this photoelectric study are located in the central region of the nebula, and their diaphragm photometry could be affected by the nebula brightness and some undetected crowding. In particular, this could be the case of the most luminous star, S294-B0.5V (MFJ No. 4), which is surrounded by several less luminous stars, more clearly visible in NIR frames. To avoid the influence of these effects, we used, for the correlation, 8 out of the 14 common stars, which are located outside the small central region marked in Fig.~\ref{fig7}. This area encloses the concentration of stars seen in the $UBVRI$ frames, and more clearly in the NIR frames. The resulting correlation between $UBV$ indices is very well defined, and provides a set of standard, MFJ-derived, $(U-B)$, $(B-V)$ and $V$ values.

These values derived from photoelectric photometry are presumably well tied to
the original Johnson system. They are, however, systematically different
from those obtained through calibration with Landolt standards.
For the 8 stars used in our correlation, we obtain
$V_{Landolt}-V_{MFJ}=-0.04 \pm 0.02$; \quad
$(U-B)_{Landolt}-(U-B)_{MFJ}=-0.07 \pm 0.03$; \quad
$(B-V)_{Landolt}-(B-V)_{MFJ}=0.01 \pm 0.03$. 
These differences between $UBV$ Johnson and Landolt systems were previously found and discussed by Delgado \& Alfaro (2000). In our case here, the ZAMS line which we use for color excess and distance determination \citep{schmidt82} is likely to be more adequate for Johnson values, and we adopt here the Johnson-derived set of standard indices. 
As for the calibration of $(V-R)$ and $(V-I)$ indices, it is obtained from the Landolt standard star measurements, following the multilinear model described by \cite{delgado98}. The uncertainties of all the standard indices obtained, given by the root mean squared deviation of the residuals, amount to 0.02, 0.03, 0.03, 0.06, and 0.03, for $V$, $(U-B)$, $(B-V)$, $(V-R)$, and $(V-I)$, respectively.

\subsection{Colour excess, distance, and age}

The standard magnitudes of all the stars detected in our $UBVRIH\alpha$ images are listed in Table~1\footnote{Table~1 is available in electronic form at the CDS via anonymous ftp to cdsarc.u-strasbg.fr or via http://cdsweb.u-strasbg.fr/cgi-bin/qcat?J/A+A/vol-number/page-number.}.
In this Table, columns (1), (2) and (3) identify the stars giving respectively, the running ID number, and the equatorial coordinates (R.A. and Dec., epoch 2000); columns (4) through (14) give the photometric indices $V$, $(U-B)$, $(B-V)$, $(V-R)$, $(V-I)$, $R$, $I$, $(R-H\alpha)$, $J$, $H$, and $K_S$, and their corresponding uncertainties in parenthesis. Column (15) gives the cluster membership assignments from optical photometry (see section 5.3). Column (16) indicates whether the source has near-IR excess. Finally, column (17) contains the identifications of the stars if it appears in the WEBDA data base\footnote{This research has made use of the WEBDA database, operated at the Institute for Astronomy of the University of Vienna.}.

\begin{figure}
   \centering
   \includegraphics[width=8.5cm, angle=0]{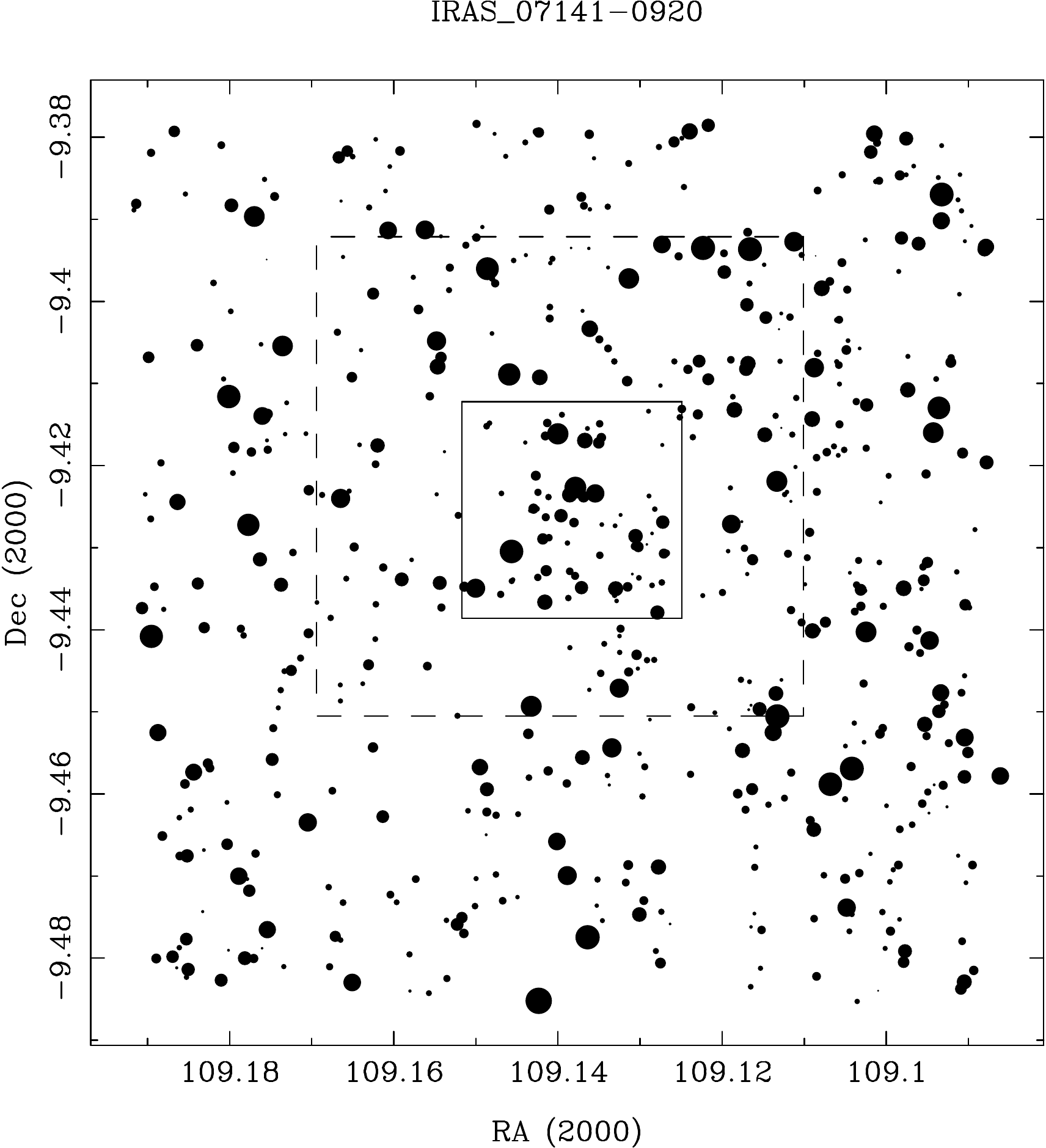}
   \caption{Schematic representation of the field covered by the $UBVRI$ observations. Only stars detected in the $V$-band are plotted. The smaller central square (solid-line) defines what we name {\em core-stars} (see text). The larger dashed-line box indicates the field of the $JHK_S$ NIR images.
	} 
	\label{fig7}%
    \end{figure}

A schematic chart of the field covered by our optical observations is plotted in Fig.~\ref{fig7}, where all stars with $V$ measurements are shown. 
In the following discussion, we will first concentrate on the stars located in the smaller square region at the center of the field, indicated in Fig.~\ref{fig7}. For simplicity, we refer to these stars as the core-stars.

%

\begin{figure}
   \centering
   \includegraphics[width=8.5cm, angle=0]{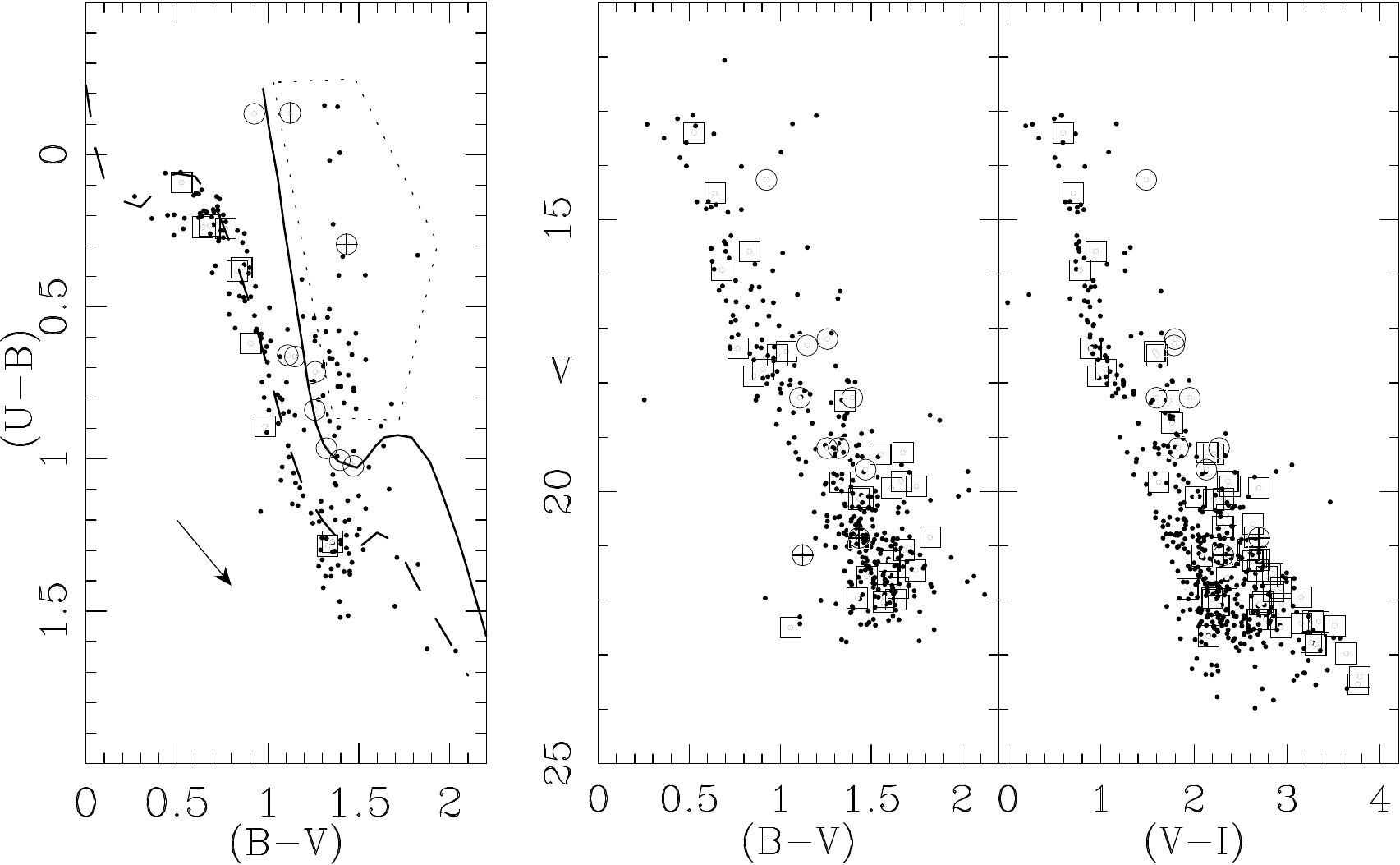}
   \caption{$(U-B)$-$(B-V)$ CC diagram and $V$-$(B-V)$, $V$-$(V-I)$ CM diagrams. The so-called core stars (see text) are shown as squares and circles. The ZAMS line is plotted twice in the CC panel: with $E(B-V)=0.1$ (dashed line, to account for the foreground field) and with E(B-V)=1.3 (solid line, to account for the sequence of reddened core-stars). The shifted line reproduces very well the sequence of reddened core-stars (circles), which we name nebula stars (see text). In both shifts the reddening slope $E(U-B)/E(B-V)=0.72$ is applied. The two crossed circles are core stars rejected as MS nebula stars from their simultaneous location in all three diagrams. The dotted line in the CC diagram encloses a region with stars possibly associated with the nebula and exhibiting $(U-B)$ excess. Note also the distinctly deeper magnitude limits as we move towards redder bands, from $U$ to $I$.
	} 
	\label{fig8}%
    \end{figure}

In Fig.~\ref{fig8}, we plot the Colour-Colour (CC) diagram $(U-B)$ vs $(B-V)$, and the Colour-Magnitude (CM) diagrams $V$ vs $(B-V)$ and $V$ vs $(V-I)$ of all the observed stars. 
Among these stars, the core-stars from Fig.~7 are represented by open circles and open squares. 
The latter appear to be part of a group of foreground stars at very low reddening.
In the CC diagram (fig~\ref{fig8}), the ZAMS line \citep{schmidt82} shifted for a color excess of E(B-V)=0.1 (dashed line) fits well this group of stars.
These foreground stars are homogeneously distributed in the observed field, and thus include some core-stars, namely, those we represent by squares.

\addtocounter{table}{1}

\begin{table}
\begin{minipage}[t]{\columnwidth}
\caption{Parameters for nebula stars}             
\label{table:2}      
\centering                          
\renewcommand{\footnoterule}{}  
\begin{tabular}{r c c c c}        
\hline\hline                 
N\footnote{ID number given in Table 1 (only available on-line).} & $ E(B-V)$ &  $E(V-I)$ & $R_V(I)$ & $DMo$ \\
\hline                        
  520 &  1.18 &  1.66 & 3.38 &  13.4 \\
  580 &  1.27 &  1.88 & 3.56 &  12.7 \\
  660 &  1.31 &  2.32 & 4.26 &  12.5 \\
  694 &  1.35 &  1.87 & 3.33 &  12.0 \\
  697 &  1.28 &  1.79 & 3.37 &  13.2 \\
  717 &  1.20 &  1.75 & 3.51 &  12.6 \\
  733 &  1.23 &  1.86 & 3.63 &  12.1 \\
  755 &  1.28 &  1.83 & 3.43 &  11.9 \\
\hline                                   
\end{tabular}
\end{minipage}
\end{table}
On the other hand, the open circles appear to be a sequence of reddened core-stars. In fact, the solid line is the previous ZAMS line shifted to a location defined by E(B-V)$\approx$1.3, with a reddening slope $\alpha$=0.72. With these parameters 
the solid line fits well the shape of the most defining part of the sequence, namely that close to late B and early A type stars. Thus, we assume that the stars represented with open circles are very likely to be reddened stars associated with the nebula. They correspond to the visible members of the young stellar cluster, better identified in the spatial distribution seen in the NIR images. We refer to them as {\em nebula-associated stars}, or just simply {\em nebula stars}. 
The two stars represented by crossed circles, which are indeed among the core-stars, are not considered in principle MS nebula stars, judging from their locations in this diagram and in the CM diagrams. They belong to a conspicuous group of stars, which could be also associated with the nebula, and showing at the same time a possible excess in the $(U-B)$ index. This is a feature associated with accretion in forming stars, as discussed by Gullbring et al. (1998, 2000). A more detailed discussion of these stars is postponed to the discussion of cluster membership.

We are therefore left with 8 stars (listed in Table~2), which we consider as the most probable nebula stars. Table~2 also provides a quantitative estimate of the colour excess $E(B-V)$ for the nebula stars, using the value of the reddening slope $\alpha$=0.72. The calculated value for the nebula illuminator, star MFJ No.4 (No. 717 in Table~2), leads to a spectral type estimate of B0.5V for this star, in good agreement with MFJ and close to the spectroscopic type estimated by \cite{russeil07}.

%

   \begin{figure}
   \centering
   \includegraphics[width=8.5cm, angle=0]{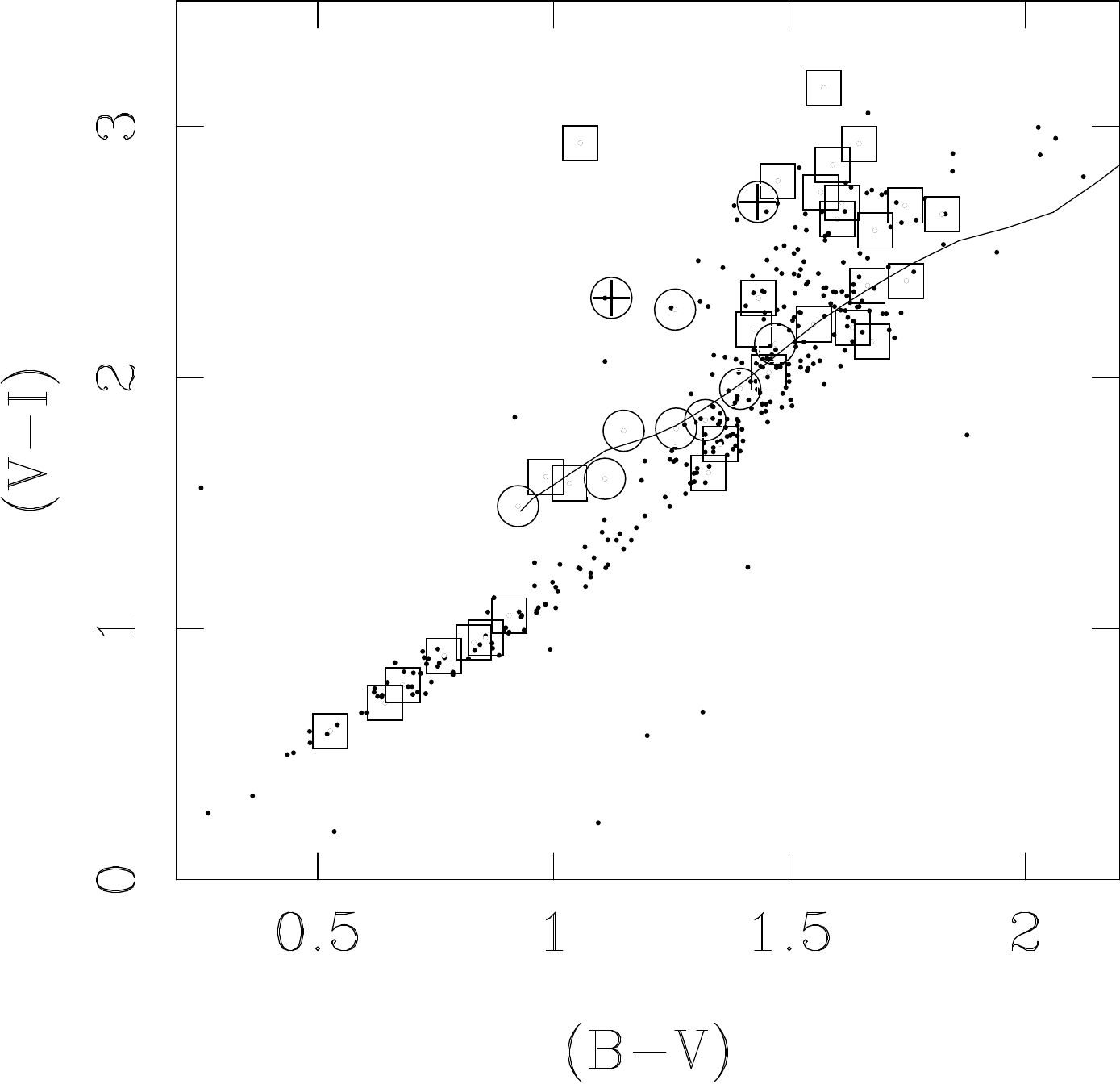}
   \caption{$(V-I)$ vs $(B-V)$ Colour-Colour plot. Symbols for core-stars and nebula stars are as in Figure 8. The ZAMS line is shifted for a color excess of $(E(B-V)$=1.26, and a reddening slope $E(V-I)/E(B-V)$=1.44.
	} 
	\label{fig9}%
    \end{figure}
%

With the values of the colour excesses listed in Table~2, a determination of the visual absorption is possible, once the visual absorption coefficient $R_V \equiv A_{V}/E(B-V)$ is obtained. Here we dispose of $JHK_S$ observations, which allow the estimate of $R_V$  for individual stars with well defined $E(B-V)$ color excess. 
For a particular star, an E(V-K) value is calculated in the V-K vs B-V diagram, by shifting the ZAMS line \citep{koornneef83} in the B-V axis with the known E(B-V), and calculating the distance to the line in the V-K axis. For our nebula stars we then obtain Rv values through the relation Rv(K)=1.12*E(V-K)/E(B-V)+0.02 \citep{fitzpatrick99}.
These values provide an average $R_V(K)=4.5 \pm 0.6$ from our NIR observations, and $R_V(K)=4.3 \pm 0.6$ with $K$ magnitudes from 2MASS. These high values of the absorption coefficient are however most probably affected by the NIR properties of the nebula stars. At least two of the nebula stars in Table~2 (520 and 660) show clear signs of NIR excess. In the presence of probable, or even suspected $K_S$-band excess, the use of the above relation is not adequate. On the other hand, a more reliable estimate of the absorption coefficient can be obtained from $(V-I)$ photometry (\citet{guetter97}; GT97 in the following). The simple relation used by these authors,  $R_{V}(I)=2.40 \times \frac{E(V-I)}{E(B-V)}$, is valid for $(V-I)$ indices in the Kron-Cousins system. The resulting Rv values are adopted here. In Fig.~\ref{fig9}, we have plotted the CC diagram $(V-I)$ vs $(B-V)$. Symbols have the same meaning as in Fig.~\ref{fig8}. The line plotted  represents the ZAMS \citep{schmidt82}, shifted to account for the calculated average value of $E(B-V)$, and a reddening slope $\frac{E(V-I)}{E(B-V)}=1.44$. We can see that this value of the slope is adequate for all nebula stars. The resulting values of $E(V-I)$ and $R_{V}(I)$ are listed in Table~2.

The values of the distance modulus obtained are also listed in Table~2. The average values $E(B-V)=1.26 \pm 0.06$, and DM=$12.5 \pm 0.5$ are obtained for the nebula stars listed in Table~2.  Even with the large dispersion, this value is distinctly smaller than the distance given by MFJ, but agrees well with the revised value of 3.2 kpc obtained by \citet{russeil07}. The larger distance obtained by MFJ is due to their adopted value for the absorption coefficient, around $R_{V}$=3.1, too small when compared to our calculated values. The presence of $R_{V}$ values both smaller and larger than those describing the average galactic extinction have been reported in the literature (see \citet{turner94,delgado98} and references therein). GT97 find a similar value, and discuss their significance as representative of the extinction law in their line-of-sight. They argue that the peculiar absorption coefficient indeed reflects the extinction law in their line of sight, rather than being related to the material physically associated to the stars. In our case, the dispersions of the averages for reddening slope and absorption coefficient are higher than theirs, partly because of larger errors in our photometry, partly because some of the peculiarities of the extinction in the region could be indeed related to specific properties of the intracluster or circumstellar material in our region. This possiblity is suggested by the larger $R_{V}(K)$ values found in our case than the ones by GT97, whereas the $E(V-I)$ excesses and $R_{V}(I)$ values are similar to theirs.

We notice that, in general, the presence in a region of $R_V$ values
higher than the standard value (R$_V$=3.1) adopted for the average
galactic extinction, might be caused by absorption due to material
associated to the observed stars, and is usually mentioned in connection
with the observation of young star clusters \citep{turner94,deWinter97}. On line with this reasoning, the total reddening in the
direction of the object would be composed of a foreground component
which can be assumed to follow a normal reddening law, characterized by
$R_V=3.1$, and a component originating in the intracluster material,
associated to the young stars, and responsible for the increase of the
R$_V$ value \citep{pandey03,sharma07,pandey08}. In the
present case, the evidences from our CC diagrams (Figs. 8 and 9)
indicate that only a small part (0.1 to 0.2 dex in $E(B-V)$) of the
reddening affecting the nebula stars is to be considered as due to
foreground reddening. Assuming that the absorption due to this part is
described by R$_V$=3.1, would have little influence on the final
distance estimate.

\subsection{Cluster membership: optical photometry assignments}

   \begin{figure}
   \centering
   \includegraphics[width=8.5cm, angle=0]{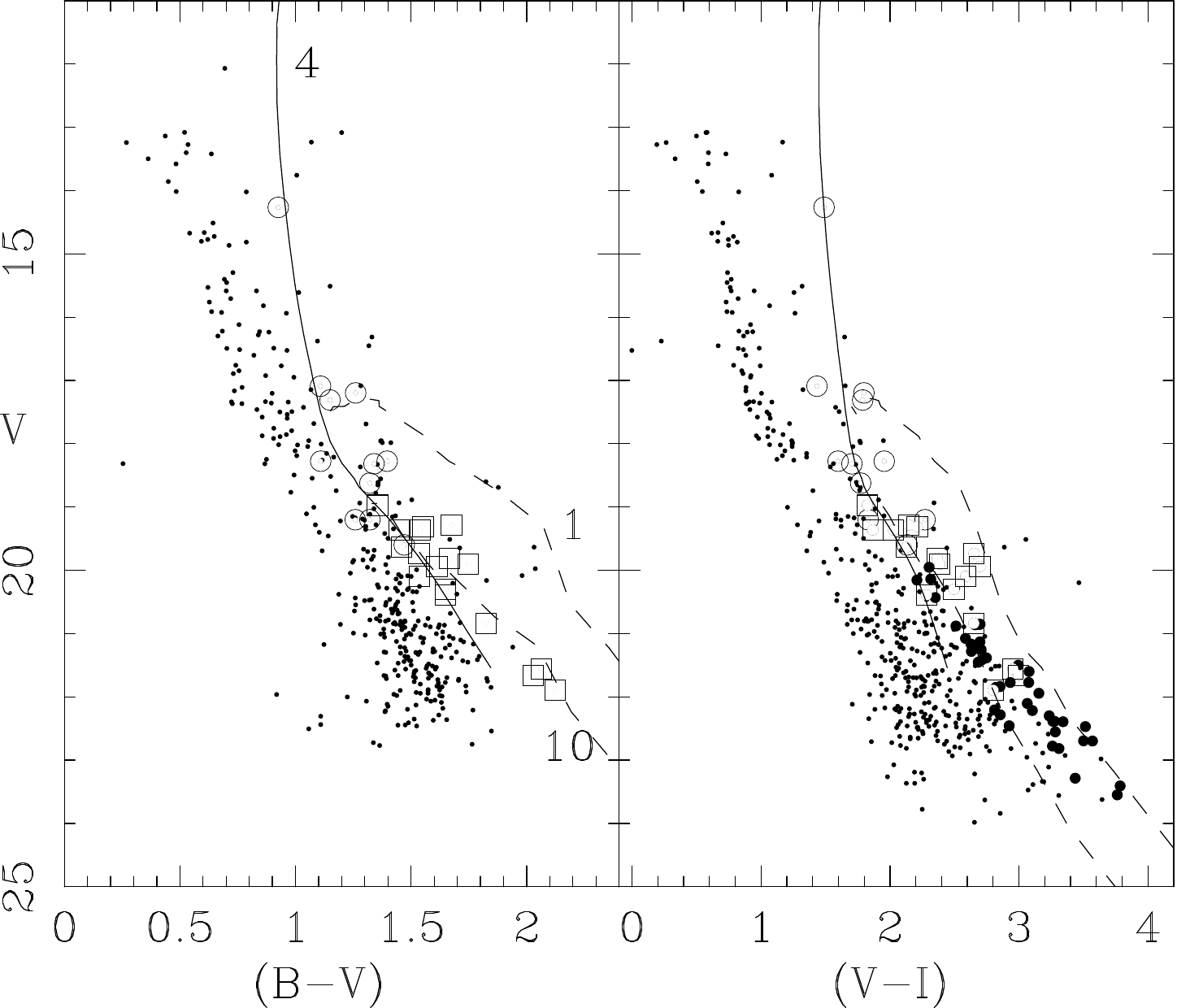}
   \caption{CM diagrams $V$ vs $(B-V)$ and  $V$ vs $(V-I)$, with indication of cluster member candidates. Circles are MS and post-MS members. They include the nebula stars in  Table~2. Large dots are PMS members assigned in the two CM diagrams $V$ vs $(V-R)$ and  $V$ vs $(V-I)$. Squares are PMS members assigned in three CM diagrams, $V$ vs $(B-V)$, $V$ vs $(V-R)$ and  $V$ vs $(V-I)$. The LogAge(yr)=6.6 ($\simeq 4$ Myr) post-MS isochrone of \citet{girardi02} is plotted (solid line), shifted to account for the colour excess and distance derived for star S294B0.5V (MFJ No.4). The \citet{palla99} isochrones for 1 and 10 Myr are also plotted (dashed lines), shifted for the average colour excess and distance modulus computed from the values of the nebula stars (see Table~2). Isochrones are labeled with their ages in Myr.
	} 
	\label{fig10}%
    \end{figure}

Using the procedures described in detail by \citet{delgado07}, we have analysed whether each star is likely to belong to the young cluster.
This membership analysis is applied to the optical $UBVRI$ observations. It consists of calculating colour excesses and distance moduli for all stars with respect to the ZAMS and PMS isochrones. A star is then considered a cluster member candidate (MS or PMS), if its colour excess, and distance modulus, with respect to any of the comparison lines, coincide (within the errors) with the corresponding average values calculated for a previously selected number of {\it bona fide} unevolved MS members. In the present calculation, the MS members are the 8 nebula stars mentioned above, which provide the values listed in Table~2, and the corresponding average value of colour excess and distance modulus of the cluster.

The reference lines used in the comparison are the ZAMS line \citep{schmidt82}, and the PMS isochrones from two different models (\citet{palla99}, and \citet{siess00}, in the following P99 and S00, respectively). When the distance modulus for a star is calculated from comparison with PMS isochrones, we obtain a PMS membership guess for it. The comparison can be performed in several CM diagrams. Some differences occur when using P99 or S00 models: the number of stars that fullfil the membership criteria are different: 58 stars turn out to be candidate PMS members with respect to P99 (assignment in 2 or 3 CM diagrams), whereas only 38 fullfil the requirements when compared to S00. 
The bulk of this difference arises in the number of stars assigned as members around the spectral types A and F. Assigned PMS members of spectral types around G and later coincide for both isochrone models. The difference is due to the fact that the S00 isochrones are more luminous (around 0.5 brighter in $M_V$ around A and F spectral types) than the P99 isochrones. As a result, also the average age of the PMS candidate members is younger when measured using the P99 isochrones. On the other hand, practically all stars assigned with respect to S00, are also assigned with respect to P99 models.
The assignments listed in Table~1 were considered with respect to either of the two models. The entries in column (16) are: 0, for no assignment; 1, for MS  members; 2, for PMS members assigned in 2 CM diagrams; and 3, for PMS members assigned in 3 CM diagrams.

The CM diagrams in Fig.~\ref{fig10} show the membership assignments obtained in the analysis. In these plots, we show the post-MS isochrone of logAge(yr)=6.6 \citep{girardi02}, shifted to account for the parameters of color excesses derived for S294B0.5V. PMS isochrones of 1 and 10 Myr from P99, shifted to account for the average parameters derived above, are also plotted. Stars indicated with larger dots in Fig.~\ref{fig10} are those with a PMS membership assignment in two CM diagrams ($V$ vs $(V-R)$ and $(V-I)$). This means \citep{delgado07} that both colour excess and distance modulus calculated with respect to some PMS isochrone between 1 and 10 Myr, coincide (within errors) with the average values obtained from the nebula stars. Similarly, stars assigned as members in three CM diagrams are marked as squares.
We detect PMS candidates down to masses around 0.7 $M_{\odot}$.
The average age value for these PMS candidates can be taken as an age estimate for the possible PMS sequence. The values obtained for stars with membership assignments in at least three CM diagrams, are 7.4$\pm$1.3 Myr for P99 isochrones, and 8.3$\pm$0.7 Myr for S00 isochrones. These values are higher than the age estimate obtained for the MS members. The significance of this estimate has yet to be assessed. If confirmed, it would indicate a sequential star formation in the region, with the most massive stars being formed last. This kind of trend is also suggested by the results on other regions of active formation (\citet{mamajek02}, \citet{sartori03}, \citet{delgado07}).

\subsection{Cluster members with near-IR excess}

\begin{figure}
   \centering
   \includegraphics[width=8.5cm, angle=0]{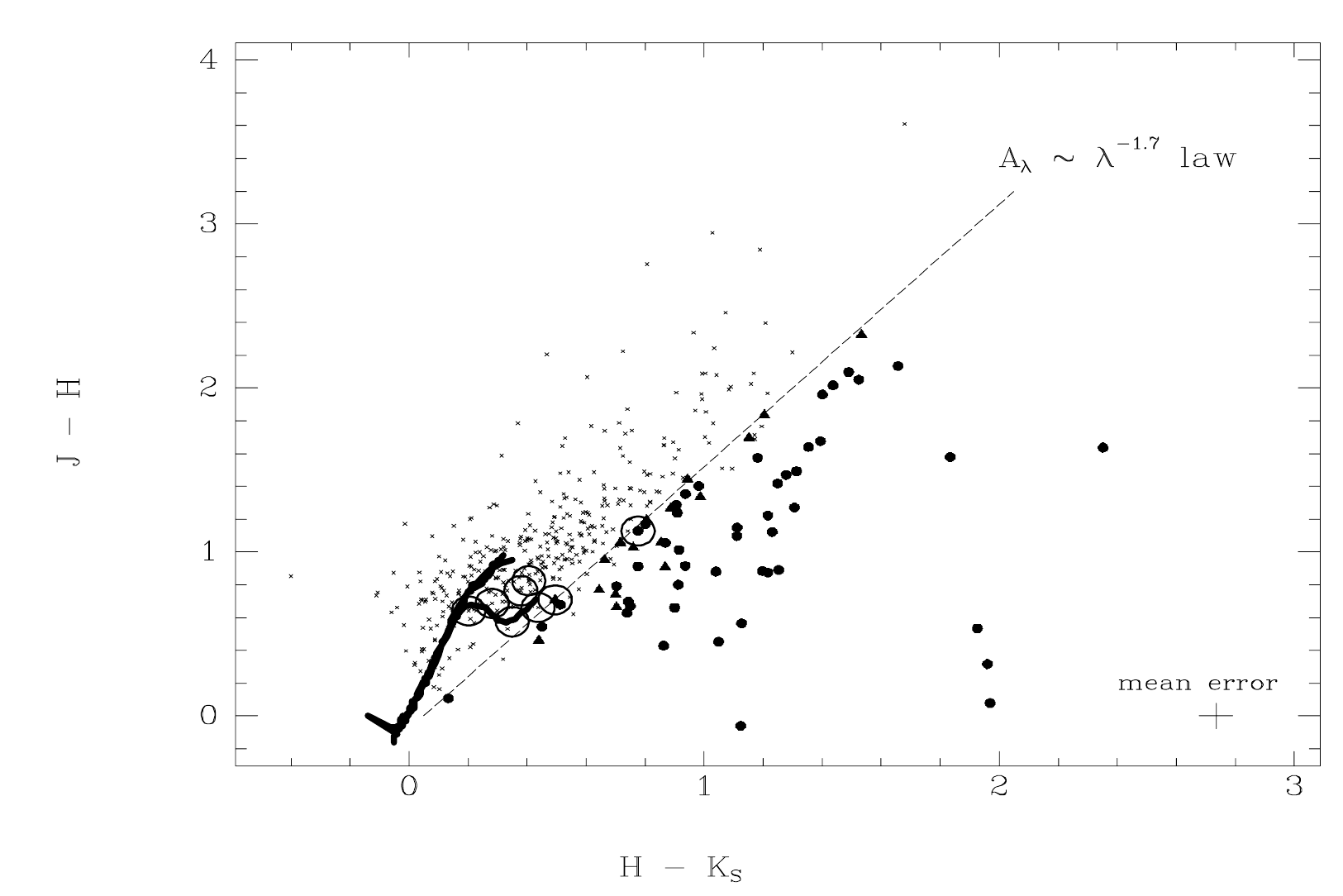}
\caption{$J-H/H-K_S$ diagram for the 437 sources with $\sigma_{Ks} < $
0.2 mag. The reddening slope of 1.6 results from applying the $A_{\lambda} \propto \lambda^{-1.7}$ parametrization of the NIR extinction law and the
reddening vector of an A0 star is drawn (dashed line). The bona-fide 
IR-excess sources (big dots) are located more than 2 $\sigma$ to the 
right of and below the reddening line and the probable IR-excess sources (triangles) have a separation from the reddening line between 1 and 2
$\sigma$ of the individual errors in the $J-H$ and $H-K_S$ indices.
Large circles mark nebula stars listed in Table~2.
}
	\label{fig-jhk}
    \end{figure}

The $J-H/H-K_S$ diagram for 437 NOTCam sources with $\sigma_{Ks} < $
0.2 mag is shown in Fig.~\ref{fig-jhk}. The loci of giant, supergiant, 
and main-sequence stars \citep{koornneef83} are indicated with bold curves. 
We have used the $A_{\lambda} \propto \lambda^{-1.7}$ parametrization of 
the NIR extinction law \citep{whittet88} to calculate a reddening slope of 
1.6 in the $J-H/H-K_S$ diagram for the NOTCam $JHK_S$ filter pass bands 
at $\lambda\lambda$ 1.247, 1.632, and 2.140 $\mu$m, respectively. The 
reddening vector for an A0 star is plotted as a dashed line. Most stars 
are located around the reddened loci of main-sequence and giant stars. 

Sources located right- and downwards of the reddening vector of an A0 
star have excess emission in the infrared, a clear sign of hot 
circumstellar dust. We define as bona-fide IR-excess sources those 
separated from the reddening vector down and right by more than 2$\sigma$ 
of the individual errors in the $(J-H)$ and $(H-K_S)$ indices (48 sources) 
and as probable IR-excess sources those with a separation between 
1$\sigma$ and 2$\sigma$ (17 sources). These are marked as big dots and
triangles, respectively, in Fig.~\ref{fig-jhk}. The 8 encircled sources 
in the figure are the so-called nebula stars discussed in Sect. 5.2.
Two of them have clear or probable near-IR excesses (IDs 520, 660) and
one (ID 733) is uncertain (see Table~2).
In Table~1 the column IREX has the value 0 for no IR-excess, 
1 for probable IR-excess and 2 for bona-fide IR-excess sources. The 
sources with excess emission in the infrared owing to hot circumstellar 
dust are candidate Young Stellar Objects (YSOs), either pre-main sequence 
(PMS) stars with circumstellar discs or protostars embedded in a more 
spherical dust distribution. We cannot separate clearly between PMS stars 
and protostars only from near-IR photometry, but because of the different
time scales of these populations, the majority of the IR-excess YSOs are probably PMS stars with discs.

There are six very red objects near the centre of the embedded cluster 
which are not detected in the $J$ band. The lower limit $(J-H)$ values 
suggest that three of these have IR excesses and are probably very 
embedded YSOs in their earliest evolutionary phase. These have been 
given the IREX indicator value 1 in Table~1.

Only 144 (272) of the 444 near-IR sources have $V$ ($I$) band counterparts, 
and these are typically the less extinguished sources with $A_V$ less than
10 mag approximately. Of the 68 IR-excess selected YSOs only 10 (30) are detected in $V$ ($I$), and of these, 2 have been designated PMS stars from 
the optical photometry fulfilling the critera of two CM diagrams (ID 412, 
473, 822). Also, 3 of the 12 optically selected MS stars have IR-excesses 
(ID 343, 520, 660). We note that our near-IR observations cover only the central $3.7' \times 3.4'$ part of the $6.5' \times 6.5'$ optical field (see Fig.~7). 

The analysis of the optical photometry finds a population of 58 PMS 
sources, of which only 2 overlap with the 68 IR-excess YSOs. This shows 
that the two photometric methods are complementary and sample different 
parts of the YSO population of this type of regions. The optically 
selected PMS stars are in general expected to be closer to the main-sequence phase 
and have less or no measureable circumstellar dust. It is also known that only about 60\% of the Classical T Tauri stars in 
Taurus-Auriga have measurable NIR excesses 
\citep[see Fig~7 in ][]{strom93}. In addition, a large fraction of the IR-excess population is not optically visible,
making a direct comparison difficult. 

The fact that 25\% of the optically selected {\it main-sequence} stars show 
IR-excesses is probably related to the fact that these are all relatively 
massive, and in theory they arrive on the main-sequence already while 
still accreting material, skipping the PMS phase. It is also possible 
that these 3 out of 12 main-sequence stars are unresolved binaries.

The IR-excess population may not be well sampled by near-IR photometry, 
however. The mid-IR ISO\footnote{Infrared Space Observatory} surveys of 
young embedded clusters demonstrated well that in all regions only about 
50\% of the IR-excess population is sampled by near-IR photometry 
\citep{kaas01}. Mid-IR photometry can better separate IR-excess from 
reddening and detect hot dust which is obscured from view. 

\subsection{The luminous protostar candidate}

\begin{figure}
   \centering
\includegraphics[width=8.5cm, angle=0]{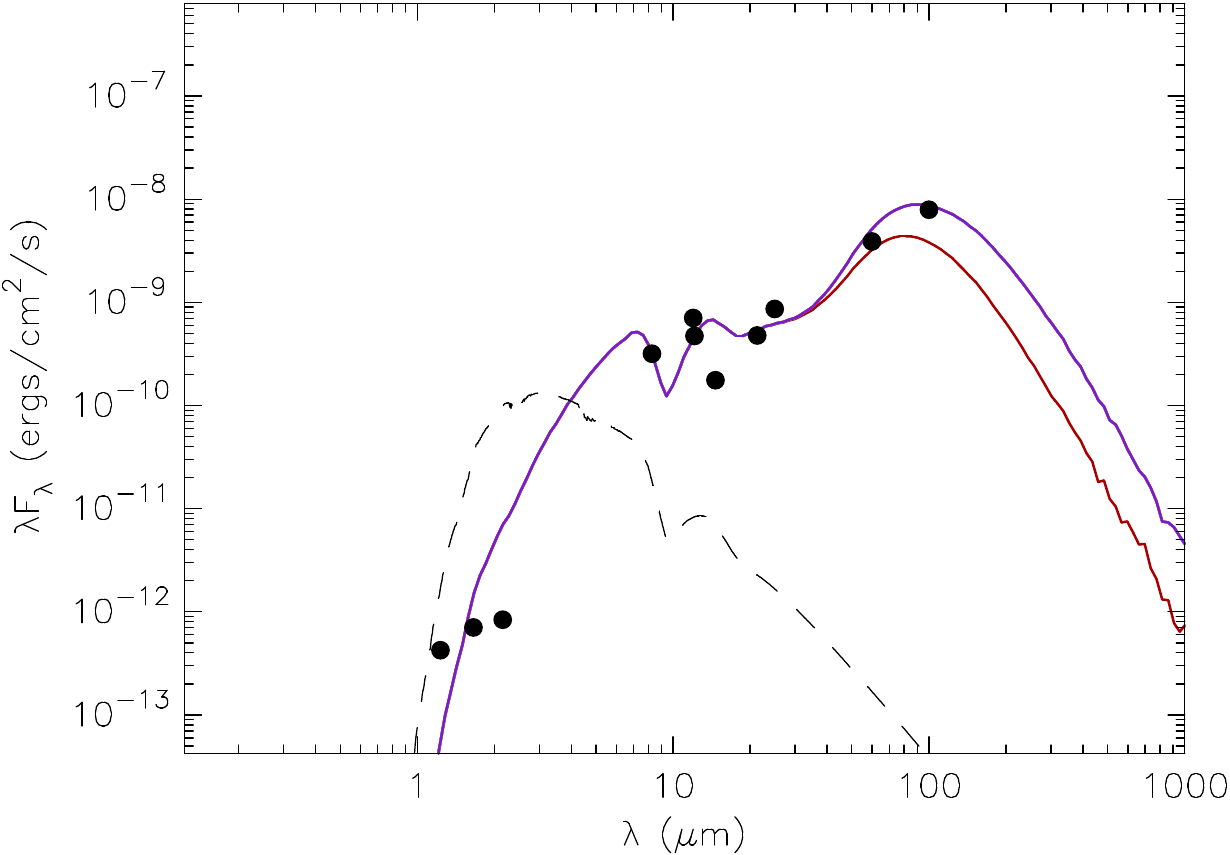}
\caption{The Spectral Energy Distribution (SED) of the luminous protostar 
candidate sampled by NOTCam $JHK_S$ band fluxes, MSX ACDE band fluxes, and 
IRAS 12, 25, 60, and 100 $\mu$m fluxes. The best fitting model found using 
the online SED fitter \citep{robitaille07} is plotted as a
continuous line that splits in two at longer wavelengths for small (lower)
and large (upper) aperture sizes. The Y-axis is scaled to a distance of 1 kpc.
}
\label{fig-sed}
\end{figure}

Searching the IRSA archive we found one mid-IR source inside our near-IR
NOTCam field in the MSX6C catalogue from the Midcourse Space Experiment. 
This is G224.1880+01.2407 which is located within $1.6''$ (with position 
angle 255 degrees) from a IR-excess source in our sample 
(ID = 1050, K$_S$ = 15.1 mag). This is less than 3$\sigma$ of the MSX6C 
positional uncertainty, and therefore we consider the source ID=1050 a possible near-IR counterpart. According to its Spectral Energy Distribution 
(SED) in the mid-IR (see Fig.\ref{fig-sed}), it is a protostar candidate. 
The SED index\footnote{The SED index is defined as $\alpha_{2-15} = 
- \frac{\log(\nu_{15}F_{\nu_{15}}) - \log (\nu_2 F_{\nu_2})}{\log 
\nu_{15} - \log \nu_2}$ where $F_{\nu}$ are the flux densities (in Jy) 
at the frequencies $\nu$ (in Hz) corresponding to the wavelengths 2 and 
15 $\mu$m.} from 2 to 15 $\mu$m is as high as $\alpha_{2-15}$ = 2.7, 
which according to the YSO classification scheme \citep{lada87,greene94}, 
though strictly valid for low mass stars only, means this is a protostar 
candidate of the type Class\,I or possibly in the transition phase from 
Class\,0 to Class\,I. The brightness of the object in the mid-IR 
(0.8755 Jy at 8.28 $\mu$m) indicates it is a luminous/massive protostar. 

Combining the NOTCam $JHK_S$ fluxes with the fluxes in the MSX bands 
$ACDE$ at $\lambda\lambda$ 8.28, 12.13, 14.65 and 21.34 $\mu$m, 
respectively, and the IRAS 12, 25, 60 and 100 $\mu$m fluxes, the SED 
is sampled over a sufficient wavelength interval to compare it to YSO 
models (see Fig.~\ref{fig-sed}). 
The online YSO SED model fitting tool provided by \cite{robitaille07} 
was used to find the best YSO model for the observed SED. The input 
ranges in interstellar extinction and distance was limited to 10~mag $< A_V <$ 
40~mag and 2.1~kpc $<$ D $<$ 4.1~kpc, respectively. The models which best 
fit the observed SED all give: a stellar mass from 8 to 12 M$_{\odot}$, 
a total luminosity of 1 to 4 $10^{3}$ L$_{\odot}$, a stellar age
of 40000 yrs or less, and an envelope accretion rate of $\sim 5 \times 
10^{-3}$ M$_{\odot}$/yr. Adjustments for aperture mismatches of the 
different wavelengths do not significantly alter the derived values.

For all best fitting models, the measured flux in the MSX~D band 
(13.5-15.9 $\mu$m) is too low. We speculate that this is caused
by very strong absorptions, such as the known feature at 15.2 $\mu$m, 
found by \citet{alexander03} to be present for the youngest and 
most embedded YSOs they surveyed with the ISOCAM CVF. The feature is 
believed to be attributed to the bending mode of CO$_2$ ice. 

\begin{figure}
   \centering
\includegraphics[width=8.5cm, angle=0]{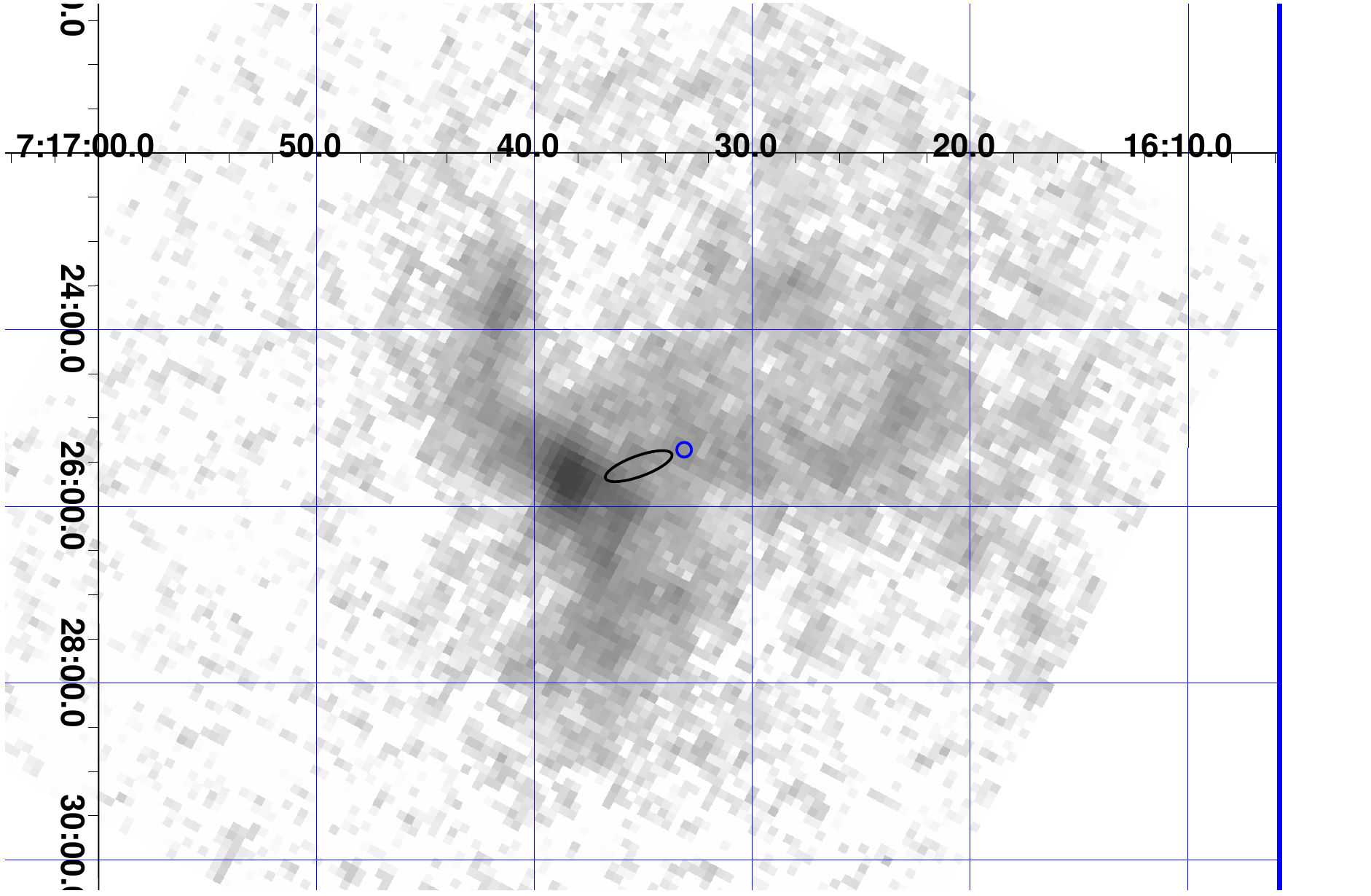}
\caption{MSX A band (8.28 $\mu$m) image with the error ellipse 
of IRAS07141-0920 taken from the IRAS point source catalogue as 
well as the location of the optical star S294-B0.5V (circle). 
The FOV is roughly as for the optical images, and the J2000 RA 
and DEC grid is shown. The intensity is in logarithmic scale.}
\label{fig-msx}
\end{figure}

An MSX A band image (6.8 $-$ 10.8 $\mu$m) covering the area of our optical
and near-IR studies is shown in Fig.~\ref{fig-msx}. The error ellipse of 
the IRAS point source is shown, and for reference, the position of the massive
main-sequence star S294-B0.5V (the optical nebula illuminator) is marked with a small circle. We note that in addition to the bright MSX point source, the 8$\mu$m 
image shows large scale diffuse emission extending as curved arms 
$\sim 2'$ or more from the point source. The diffuse emission is strong 
at 8 $\mu$m, but barely distinguished from the noise in the other MSX 
bands. Its main origin is likely the emission features at 7.7 and 8.6 
$\mu$m, two of the emission bands of polycyclic aromatic hydrocarbons 
(PAHs). 
As shown in Fig.~\ref{ha_msx_mem3}, the curved filaments in emission at 8 $\mu$m trace well the borders of the $H\alpha$ emission.
This resembles very much the findings of \citet{deharveng03} 
for the geometrically simpler sources Sh 217 and Sh 219, two spherically 
symmetric HII regions surrounded by annular photodissociation regions 
traced by both 21cm HI and 8$\mu$m PAH emission. In our case the 
morphology is more complex.


%

\subsection{PMS membership and $H\alpha$ emission}

\begin{figure}
   \centering
   \includegraphics[width=8.5cm, angle=0]{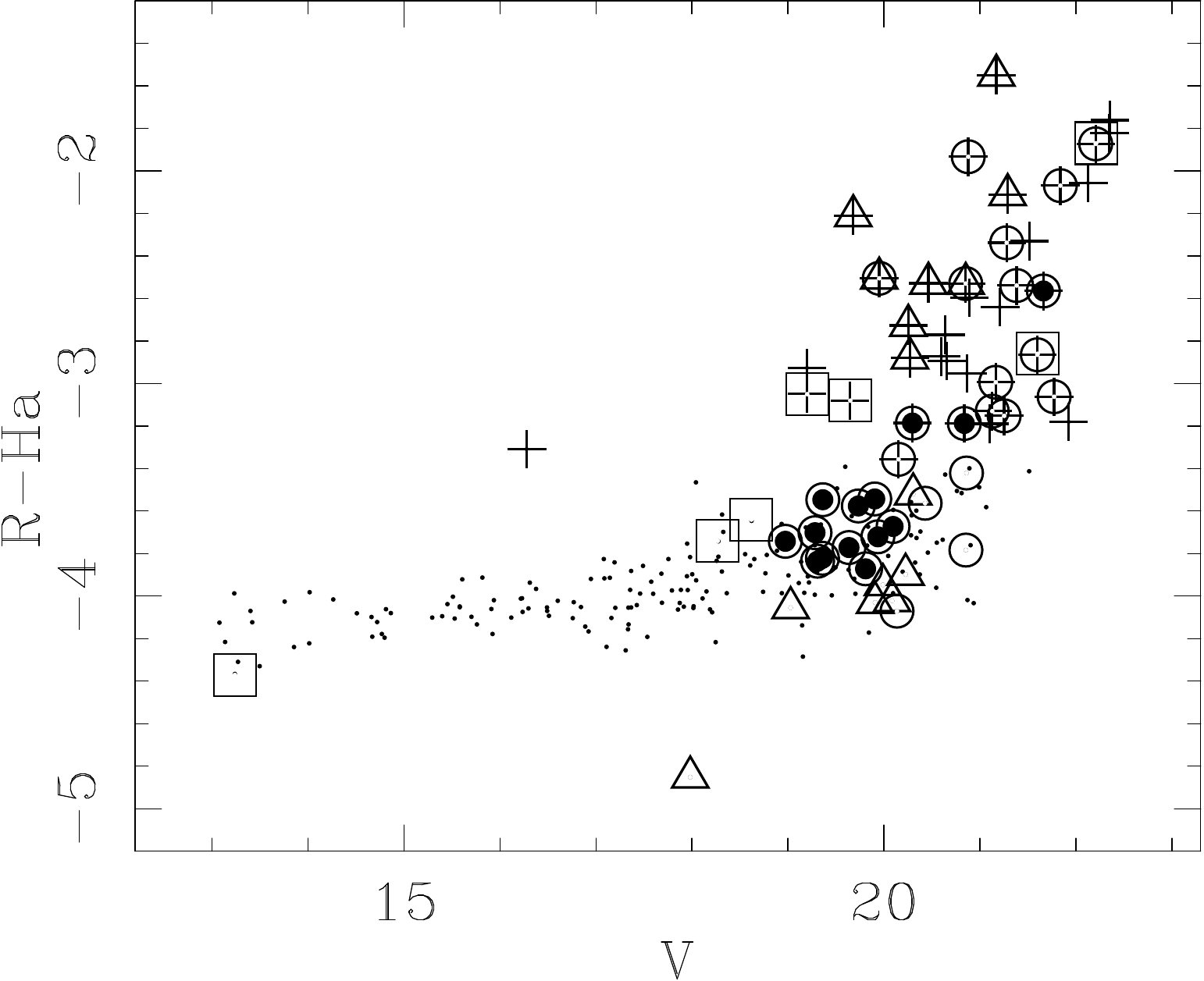}
   \caption{Plot of the instrumental index $(R-H\alpha)$ versus $V$ magnitude. Small dots represent all stars in our sample with valid $R$ and $H\alpha$ measurements. Different symbols are used as follows: squares, for stars with NIR excess; crosses, for stars with possible emission in $H\alpha$ ($(R-H\alpha) > -3.4$); open circles, for PMS candidate members, with respect to P99 or S00 models, selected in 2 CM diagrams; closed circles, PMS candidate members selected in 3 CM diagrams; triangles, for stars with possible $(U-B)$ excess.
	} 
	\label{figRHa}%
    \end{figure}

The combined evidence from all photometric diagrams shows indications of several stars associated to the cluster, which are not detectable as candidate members at optical wavelengths. We consider here spectral features which could be present in several stars in our sample, and have been identified in the literature as spectral signatures characteristic for PMS stars. 
The basic feature in this context is the presence of $H\alpha$ in emission, typical in the spectra of low mass forming stars of the TTauri type, both in classical (CTT) and weak-lined (WTT) stars, as well as in higher mass forming stars of the HAeBe type. The value of the $R-H\alpha$ index has been used in previous studies of young clusters to describe this feature, although its value may be reflecting spectral properties other than those due to the PMS nature of the stars (\cite{sung97}, \cite{sung00}, \cite{sharma07}). For our stars this index, calculated as the difference between instrumental magnitudes, is listed in column 11 of Table 1, and plotted versus $V$ magnitude in Fig.~\ref{figRHa}. Different symbols are used to distinguish stars which show different features or spectral signatures of interest: possible $H\alpha$ emission ($R-H\alpha >$ -3.4 in the plot), PMS candidates from our $UBVRI$ photometry, NIR excess, as described in Section 5.4, and probable excess in $U-B$ (see Sect 5.2 and Fig. 8).


In their spectroscopic study of the PMS population in the 4~Myr old cluster Tr~37, \citet{sicilia05} find 40\% of stars with signs of accretion, a percentage that decreases to almost zero in the 10~Myr old cluster NGC~7160. Furthermore, the percentage of these accreting stars with signs of NIR excess, 50\% in Tr~37, drastically decreases with increasing age. The corresponding numbers from our photometric analysis would be: 16 stars with $(R-H\alpha) > -3.4$ out of 58 stars selected as optical PMS candidates (27\%), and 4 stars among these 16 (25\%) with NIR excess. Considering $H\alpha$ emission as a possible tracer of accretion processes, and given the age of our optically selected PMS sequence (around 8 Myr) the corresponding percentages in our sample are consistent with those results. 

Figure~\ref{figRHa} indicate that the presence of possible $(U-B)$ excess coincides frequently with the indication of $H\alpha$ emission, but it seems to exclude PMS candidates selected in 2 or 3 optical CM diagrams. In particular, those of earlier spectral type (around AF: stars marked with both an open and a closed circle), most do not show any sign of circumstellar activity or ongoing accretion, whereas several fainter PMS candidates, below spectral type G and later (marked with only open circles), do show signs of $H\alpha$ emission, but only two of them possible $(U-B)$ excess. This mutual exclusion of PMS nature (inferred from optical CM diagrams) and $(U-B)$ excess might have two possible causes. Stars with $(U-B)$ excesses might have other spectral anomalies, such as $(B-V)$ indices with some excess, and possible $H\alpha$ emission. These features would prevent them from being selected as optical PMS candidates by our CM diagrams method. On the other hand, and more interestingly, 
the disappearance of $(U-B)$ excesses as we move to earlier type stars could be due to magnetospheric accretion stopping operating in HAeBe stars due to weaker or absent magnetic fields \citep{wade07} when compared to magnetic fields in lower-mass PMS stars.

In general, the progressive disappearance in our PMS sample of spectral features, as $H\alpha$ emission, and even $(U-B)$ excess, as we move to earlier spectral types, is a finding on line with arguments in favour of a faster dissipation of accreting discs and envelopes in stars of higher masses \citep{hernandez05}. In fact, a decrease in number and intensity of emission features in stars of spectral type F and earlier has been found in comprehensive analyses of the spectral types of PMS stars \citep{hernandez04}. Furthermore, the results of \citet{manoj06}, relative to HAeBe stars, show a rapid decrease of $H\alpha$ emission with age, but more interestingly, no F-type star in their sample exhibits an equivalent width of $H\alpha$ in emission above 10, independently of age. In this same context, \citet{hernandez06} have found that the PMS population in Orion, between ages of 5 and 10 Myr, show signs of residual debris discs, without excess in the IRAC bands and without emission lines in their optical spectra, but with a varying degree of 24 $\mu$m excess.

\section{Discussion}

\subsection{Spatial distribution}

A total of 133 cluster member candidates were found in the 
previous section using CM and CC diagrams based on $UBVRIJHK_S$ 
photometry. From the optical CM diagrams using the method of
\citet{delgado07} we find 12 main-sequence stars and 58 PMS 
stars. Using IR-excess as a youth and member criterium we find 
68 PMS members from the $J-H/H-K_S$ diagram. Only 5 of these 
coincide with the optically selected members.
In addition, we find 43 sources with strong 
$H\alpha$ emission using the $R-H\alpha$ index. Of these, 21
coincide with members selected from optical CM diagrams or 
with NIR excess. The total number of cluster member candidates 
found in this study is thus 155.

\begin{figure}
   \centering
\includegraphics[width=8.5cm, angle=0]{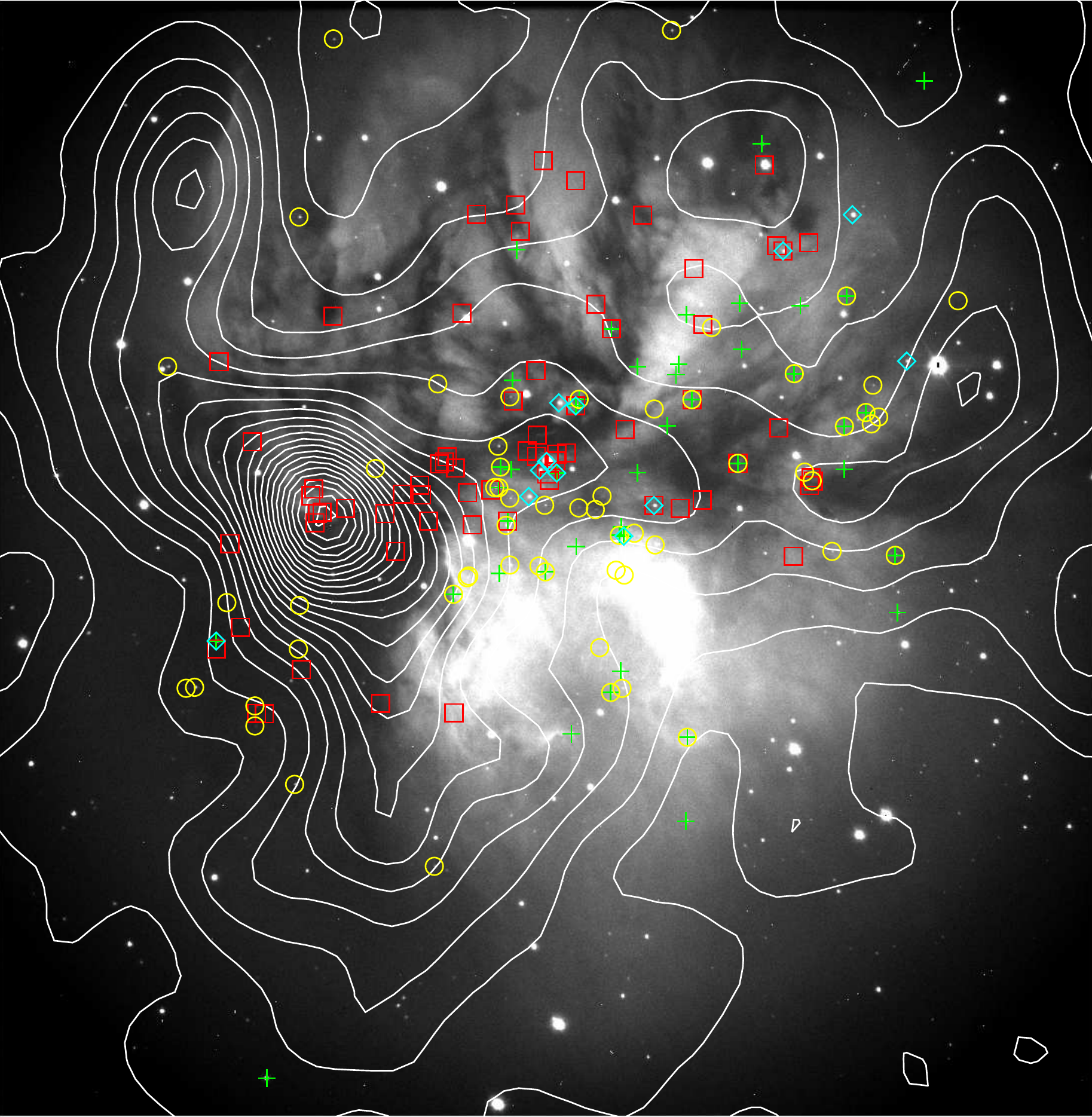}
\caption{The spatial distribution of the cluster members 
overlaid on the $H\alpha$ image (grayscale). The white contour lines represent the MSX 8 $\mu$m isophote emission.  The positions of main-sequence 
stars (blue diamonds), PMS stars selected from optical CM diagrams 
(yellow circles), IR-excess stars selected from the $J-H/H-K_S$
diagram (red squares), and $H\alpha$ emission sources selected 
from the $(R-H\alpha)$ index (green crosses) are shown. The massive MSX protostar candidate is 
located at the strongest peak
of the 8 $\mu$m emission. The spatial distribution of the IR-excess sources is strongly sub-clustered. }
\label{ha_msx_mem3}
\end{figure}

The spatial distribution of the cluster members can give valuable
information about cluster evolution and star formation scenario. 
In Fig.~\ref{ha_msx_mem3} we show the location of the cluster member 
candidates with different symbols according to selection method (see
figure caption for details) overlaid on the $H\alpha$ image. While 
it is clear from Fig.~\ref{fig6} that there are mainly two stellar 
density peaks at 2 $\mu$m, referred to as the eastern and western 
clusters, the spatial distribution of the cluster member candidates, 
as selected by our previously described methods, gives a more nuanced
impression. The density of cluster members is still highest at the
location of the eastern and western cluster, but about half of the 
members are extended both towards SE and NW, tracing the dense middle 
lane of the region seen in absorption on the $H\alpha$ image. On a 
large scale, star formation seems to be going on along a filament or 
flattened density structure along a SE-NW oriented ridge. 

As many as 8 of the 12 optically selected MS cluster members are 
located in the area of the strongest (western) stellar density peak seen at 
2.12 $\mu$m (see Fig.~\ref{fig6}) around the source S294-B0.5V. 
Three are located in an area about $1.5'$ to the NW, and the last 
one about $1.5'$ to the NE. We note that there is only one MS member
found towards the SE.

The optically selected PMS cluster members are mainly distributed 
in roughly three loosely clustered areas: one around the exciting 
source S294-B0.5V, one about $1.5'$ to the NW, and one about $1.5'$ 
to the SE. 

The IR-excess selected YSOs have a spatial distribution mainly across 
the NW-SE diagonal of the $3.7' \times 3.4' JHK_S$ image. In addition
to the more scattered component of IR-excess source distribution, 
about half of them are strongly clustered. Around the source S294-B0.5V 
and coinciding with the strongest stellar density peak of Fig.~\ref{fig6} 
as many as 8 IR-excess sources are inside $\sim 16''$ (see the zoom-up 
in Fig.~\ref{fig5} for high spatial resolution). Towards the eastern
cluster, there is clustering of IR-excess sources in a dense part of
the cloud practically void of optical sources. 
The projected separations between all the near-IR excess sources have a
distribution with one peak at about 15 arcsec and another at 90 arcsec. The 
latter reflects the extent of the most active region of star formation,
about 1.4 pc for an adopted distance of 3.2 kpc. The peak at 15 arcsec or
0.23 pc reflects the typical size of the smallest stellar groups. This is 
in good agreement with the smallest clustering scale of Class\,II sources 
(0.25 pc) found in the nearby Serpens star forming region \citep{kaas04}.
In the easternmost of these groups resides the luminous MSX source 
G224.1880+01.2407, a massive protostellar candidate, the youngest of
the members. 

Whether the groups of the embedded eastern cluster are younger than 
the members in the western cluster around S294-B0.5V is not clear. 
The clustering of IR-excess sources is strong around S294-B0.5V as well. The 
age difference between the two most massive stars is clear, however. 
One is already on the main sequence, while the other one is still in 
its protostellar phase. 
It is also quite likely that the median age of the IR excess 
sources is less than the median age of the optically selected PMS sources,
and for comparison the Serpens Class\,II sources with a similar clustering 
scale have an age estimate of about 2 Myr \citep{kaas04}.
Kinematic data may be needed in order to better understand the nature of the sub-clustering processes observed in this star forming region.

\subsection{A possible star formation scenario for Sh2-294}

The MSX A-band image shows diffuse emission along curved filaments that trace
well the borders of the $H\alpha$ emission (see Fig.~\ref{ha_msx_mem3}). If
the 8 $\mu$m emission is mainly due to PAHs, as suggested in Sect. 5.5., this
spatial distribution can be interpreted as a photodissociation region (PDR)
forming a shell around the optical H II region, in a similar fashion as for
Sh2-217 and Sh2-219 \citep{deharveng03}, although the morphology is less
symmetrical for Sh2-294. The bright MSX point source is observed in the direction
of the eastern edge of the PDR region, only $\Delta \alpha=24.3''$ and $\Delta
\delta=4''$ offset from the eastern and strongest of the two peaks in the
1.46 GHz VLA map \citep{fich93}. The IRAS colours indicate an ultra-compact
H II region, and the quoted $7'$ diameter of the VLA source \citep{fich93}
is understood to include the free-free emission from the larger and optically
visible H II region centred on the western radio peak close to S294-B0.5V.

The appearance in the near-IR images is that of a double cluster with the eastern
cluster containing the MSX protostar, probably responsible for the ultra-compact
H II region, and the western cluster centred on the massive star S294-B0.5V,
responsible for the optical H II region. The fact that the mid-IR emission traces
so well the borders of the optical HII region strongly argues in favour of a
physical connection and against a chance alignment of two clusters at different
distances. Their projected separation is small, but the eastern, embedded cluster
must be towards the front side of the optical H II region. Its shape formed by
the sub-clustered IR-excess sources traces the east-west filament (8$\mu$m emission
and $H\alpha$ absorption). The spatial distribution of the bulk of the cluster
members, the lower mass stars, is not distinctly bimodal, however, and the
optically selected PMS stars (with typical ages of 7-8 Myr) are in many areas
spatially mixed with the IR-excess sources (most of which are not seen in the
optical, are sub-clustered, and probably much younger). This can be partly due
to projection and spatially variable cloud extinction.

The on-going massive star formation, traced by the MSX protostar
G224.1880+01.2407, seems to take place in the PDR. It may have been triggered by the optical H II region excited by S294-B0.5V, estimated to have spent about 4~Myr on the main-sequence. 
If this is the case, the age difference between the two generations is
probably quite small. For the bulk of the cluster members, the low mass
stars, it is hard to argue for any distinct age difference between the
eastern and western part, although there is probably a difference 
between the median age of the IR excess sources and that of the optically 
selected PMS stars.
In order to map the molecular cloud and get more insight into the star formation scenario for Sh2-294, follow-up observations in the mm and radio wavelengths are needed.

\section{Summary}

We present an optical and near-infrared imaging and photometric study of the HII region Sh2-294 (S294) associated with the IRAS source PSC~07141-0920. 

High-resolution images in the $UBVRI$ and $H\alpha$ bands reveal an emission nebula with very rich morphological details. The nebula is composed of ultraviolet sctattered light and of H$\alpha$ emission with local morphological features conveying an impression of a complex web of streams, flows, shocks, turbulence, cavities, a region of great dynamical complexity.
Contrasting with the bright parts of the nebula, opaque patches are seen. These dark cloud regions appear to be in the process of being shredded or blown apart by stellar winds, jets or outflows phenomena. They appear elongated, filamentary, wispy, clumpy or knotty.

Our optical photometry confirms that the illuminator of the nebula is likely to be star No. 4 of \cite{moffat79}, with a spectral type of B0.5V, but located at a distance of about 3.2 kpc, as recently revised by \cite{russeil07}.

Our high-resolution near-IR images reveal an embedded cluster, extending for about 2~pc and exhibiting sub-clustering: a denser, more condensed sub-cluster (western sub-cluster) surrounding the high-mass B0.5V illuminator star; and a more embedded sub-cluster (eastern sub-cluster) located towards the eastern dark part of the nebula and including the luminous MSX source G224.1880+01.2407, a massive protostellar candidate that could be the origin of jets and extended features seen at 2.12 $\mu$m.

This double cluster appears to be clearing the remaining molecular material of the parent cloud, creating patches of lower extinction and allowing some of the least reddened members to be detected in the optical images.

155 stars are considered as cluster members, 12 of them, MS members. The remaining are PMS candidates, inferred from three different types of indicators: membership assignment from comparison with PMS isochrones, $H\alpha$ emission deduced from the $(R-H\alpha)$ index, and presence of NIR excess. 
About half of the IR-excess sources are spatially distributed in 
sub-clusters with typical sizes of about 15 arcsec or 0.23 pc. These are
probably younger than the optically selected PMS stars.
The age estimate for the massive MS member is about 4 Myr and the average age of the optically selected PMS members around 7-8 Myr. 
Our results support time scale estimates for the duration of accretion disks around 10\,Myr for late G-type stars and later, and a remarkable decrease of this duration towards earlier spectral types.

We provide a catalog containing the photometry of all 1344 stars detected at least in one of the $UBVRIJHK_S$ bands. Cluster membership is also included.

\begin{acknowledgements}
JLY thanks the Instituto de Astrof\'\i sica de Andaluc\'\i a for hosting him during his sabbatical leave.
This work has been partly supported by the Portuguese Funda\c{c}\~ao
para a Ci\^encia e Tecnologia (FCT).
This work has been supported by the Spanish MEC through grant
AYA2004-05395 and by the Consejer\'\i a~ de Educaci\'on y Ciencia de la Junta de Andaluc\'\i a, through TIC\,101.
Part of the data presented here have been taken using ALFOSC, which
is owned by the Instituto de Astrofisica de Andalucia (IAA) and
operated at the Nordic Optical Telescope under agreement between IAA
and the NBIfAFG of the Astronomical Observatory of Copenhagen.
We made use of the NASA ADS Abstract Service and of the WEBDA data base, 
developed by Jean-Claude Mermilliod at the Laboratory of Astrophysics of the EPFL (Switzerland), and further developed and maintained by Ernst Paunzen at the Institute of Astronomy of the University of Vienna (Austria). 
This publication made use of data products from the Two Micron All Sky Survey, which is a joint project of the University of Massachusetts and the Infrared Processing and Analysis Center/California Institute of Technology, funded by the National Aeronautics and Space Administration and the National Science Foundation. 
This research made use of data products from the Midcourse Space
Experiment. Processing of the data was funded by the Ballistic Missile
Defense Organization with additional support from NASA Office of Space
Science. 
This research made use of the NASA/ IPAC Infrared Science Archive, which is operated by the Jet Propulsion Laboratory, California Institute of Technology, under contract with the National Aeronautics and Space Administration.
This research also made use of the
SIMBAD database, operated at CDS, Strasbourg, France, as well as SAOImage
DS9, developed by the Smithsonian Astrophysical Observatory.

\end{acknowledgements}

\end{document}